Review Article

# The Emergence of Pico-Kelvin Physics

Xuzong Chen, Bo Fan,

Institute of Quantum Electronics, Department of Electronics, School of Electronics Engineering and Computer Science, Peking University, Beijing 100871, China

E-mail: xuzongchen@pku.edu.cn

**Abstract:**

The frontier of low-temperature physics has advanced to the mid pico-Kelvin (pK) regime but progress has come to a halt because of the problem of gravity. Ultra cold atoms must be confined in some type of potential energy well: if the depth of the well is less than the energy an atom gains by falling through it, the atom escapes. This article reviews ultra cold atom research, emphasizing the advances that carried the low temperature frontier to 450 pico-Kelvin. We review micro gravity methods for overcoming the gravitation limit to achieve further low temperature using free fall techniques such as a drop tower, sounding rocket, parabolic flight plane and the Space Station. We describe two techniques that give promise of further advance—an atom chip and an all-optical trap—and present recent experimental results. Basic research in new regimes of observation has generally led to scientific discoveries and new technologies that benefit society. We expect this to be the case as the low temperature frontier advances and we propose some new opportunities for research.

1. Introduction

Various temperature regimes correspond to different energy scales with distinctive classes of phenomena. Consequently, the achievement of increasingly lower temperatures has been a continuing goal for physicists. Superconductivity [1] and superfluidity [2] were discovered at temperatures near 1 K. The invention of laser cooling opened the way to the microkelvin regime [3–10] and a new world of atomic quantum fluids including Bose-Einstein condensates (BECs) [11–14] and degenerate Fermi gases [15–16]. Evaporative cooling further reduced the temperature of an atomic gas to the nanokelvin regime [11–14]. Loading degenerate atoms at nanokelvin temperatures into optical lattices made quantum simulations possible [17–31], including the phase transition from a superfluid to the Mott state [19], the Handel model topological phase transition [27], and quantum dynamic phase transitions [30–31]. Low temperatures have also facilitated the application of new technologies. Cryogenic technology at a few Kelvin enabled nuclear magnetic resonance imaging [32], superconducting quantum interference device (SQUID) for high-sensitivity detection [33], and magnetic train suspension [34]. Atomic interferometry can be used to detect groundwater through gravitational anomalies [35]. Cold atoms at microkelvin temperatures enabled the creation of atomic fountain clocks with an uncertainty of one second per 30 million years [36] and can be used in global positioning and communication systems. Optical lattice clocks operating with atoms in optical lattices, such as the strontium light clock [37], can achieve an uncertainty of one second per 10 billion years, paving the way to new applications in precision frequency and length metrology.

Moving from nanokelvin to picokelvin temperatures presents a new challenge. Because laser cooling is the basis of achieving picokelvin temperatures, we briefly review the laser cooling techniques that brought us to today's frontier:

- Reducing the speed of atoms moving along the axis of an atomic beam by the retarding force of light by William D Phillips et al. [3–4, 10]
- Three-dimensional cooling of an atomic gas through the creation of an "optical molasses" (Chu, Ashkin [5], [8])
- Trapping and cooling atoms using the magneto-optical trap (MOT) (Pritchard et al. [6])
- Cooling to the microkelvin regime by "Sisyphus cooling" (Cohen-Tannoudji and Chu [7–9])
- Evaporative cooling (Hess [38–39])
- Discovery of BECs in 1995 using evaporative cooling and subsequent cooling into the nanokelvin regime by Chu and Wieman and by Ketterle [11–14]
- Achievement of 0.45 pK temperatures with magnetic field gradients to counter gravity and then using final cooling by adiabatic expansion of the trap by Ketterle [40]

The concept of temperature implies thermal equilibrium; gaseous systems require the energy distribution for the particles to be the same in all directions. For a real atomic gas system, the idea of "effective temperature" is introduced in some cases, defined by the width of the atomic velocity distribution in the system along a particular axis. This velocity may be much lower than the average velocity of the atoms in the system. For example, Raman cooling [41–42] and velocity-selective coherence population trapping [43] can cause some atoms in the system or atoms moving in one direction to obtain nanokelvin or picokelvin temperatures. The delta-kick cooling (DKC) approach is an effective two-dimensional cooling method [44]. In 2015, Kasevich's team used the transverse DKC approach in an atomic interferometer experiment to reduce the effective temperature corresponding to the transverse velocity of the launched atomic sample to 40 pK but retained the temperature of 1 nK in the longitudinal direction [45].

Obtaining picokelvin and femtokelvin temperatures in three-dimensional systems naturally generated ideas for eliminating the effects of gravity by cooling atoms in an apparatus under free-fall conditions, termed microgravity.

In 2006, Ertmer and Rasel of Germany used a drop tower to conduct Rb atomic cooling experiments on atom chips in the Quantus project. (Quantus is the abbreviation of *Quantengase unter Schwerelosigkeit*, and "atom chip" is the term adopted by the atomic community for a miniaturized magnetic atom trap generated by specially designed surface circuits to trap atoms).

In 2010, they achieved Bose-Einstein condensation (BEC) using the drop tower (120 meters high); in the following year, they reached nanokelvin temperatures [47].

In 2011, Bouyer's team in France used parabolic aircraft flight to attain cooling Rb using all-optical techniques and demonstrating atomic interference in a microgravity environment [48].

In 2013, Rasel's team at University of Hannover carried out atomic BEC interference experiments using an atom chip and a drop tower. In 2017, the same team started the MAIUS-1 project that used a sounding-rocket to provide 6 min of microgravity time and carried out Rb atomic BEC testing with an atom chip. The objective was to obtain ultracold atoms with temperatures in the picokelvin regime [52–53], and they demonstrated nanokelvin temperatures in practical experiments.

In 2015, NASA's Jet Propulsion Laboratory (JPL) team proposed to conduct Rb and K cooling experiments on the International Space Station (ISS) with the goal of obtaining an ultracold atomic gas of 100 pK [50–51]. In 2018, the United States launched the Cold Atom Laboratory (CAL) experimental module and carried out Rb and K cooling experiments on the ISS [54-56].

In 2018, Peking University demonstrated a scheme based on all-optical dipole traps, referred to as two-stage cooling (TSC) [57–60]. The TSC scheme was proposed in 2013 and is planned to be implemented on the Chinese Space Station (expected to launch in 2022), with the target temperature of below 100 pK. In 2018, a test of the TSC scheme was performed on the ground [60].

This review article will focus on the theoretical and experimental progress in the realization of picokelvin temperatures. Section 2 presents a summary of the principles of accomplishing this objective. In Section 3, the achievement of three-dimensional picokelvin temperatures by Ketterle's team and two-dimensional picokelvin temperatures by the Kasevich team is described. Section 4 describes the QUANTUS experiments of Rasel's team using a drop tower, the MAIUS-1 experiments that employed a rocket, and, finally, the CAL experiments of the JPL team using the ISS. Section 5 reviews the cold atom experiments of Bouyer's team using parabolic aircraft flight and introduces the new scheme of BEC experiments by the Peking University team using the Chinese Space Station. Section 6 describes prospects for research in science and technology that would be created by achieving the picokelvin temperature regime.

## 2. Basic Principle of Realizing picokelvin

The theory [61] states that for an $N$ number of neutral gas atoms trapped in a simple harmonic trap with frequency of $\omega_{ho}$, if the temperature $T$ of the atomic gas is lowered below the critical temperature $T_c$ in some way, with the number of atoms condensed in the ground state $N_0$, then the number of atoms in the excited state (i.e., number of thermal atoms in the background) is:

$$N - N_0 = \zeta(3)\left(\frac{k_B T}{\hbar \omega_{ho}}\right)^3 \qquad (1),$$

where $k_B$ is Boltzmann's constant, $\hbar$ is Plank's constant divided by $2\pi$, $\omega_{ho} = (\omega_x \omega_y \omega_z)^{1/3}$ is the geometric average of the frequency of the harmonic trap, and $\zeta(n)$ is the Riemann Zeta function, $\zeta(3) = 1.064$. When the number of atoms in the condensed state $N_0 \to 0$, the corresponding temperature is the critical temperature $T_c$, also known as the phase transition temperature.

$$k_B T_c = \hbar\omega_{ho}\left(\frac{N}{\zeta(3)}\right)^{\frac{1}{3}} \approx 0.94\hbar\omega_{ho}N^{1/3} \tag{2}$$

It can be seen from Eqn. (2) that the number of atoms *N* and the frequency of trap $\omega_{ho}$ determine the critical temperature $T_c$ of the Bose condensate bound in the trap. In practice, it is necessary to maintain a certain number of atoms, generally *N* between $10^4$ and $10^6$, to obtain a high signal-to-noise ratio (SNR) and high phase-space density for the experimental measurements. In order to obtain a lower temperature degenerate gas, it is necessary to reduce the trap frequency (i.e., the geometric average frequency of the trap): $\omega_{ho} = (\omega_x\omega_y\omega_z)^{1/3}$.

In the trap, the relationship between the temperature *T* of the background thermal atom and the number of thermal atoms, $N - N_0$, is as follows:

$$\frac{T}{T_c} = \left(\frac{N-N_0}{N}\right)^{1/3} \tag{3}$$

From Eqn. (1), the temperature of the background atom in the trap can be obtained as follows:

$$T = 0.94\frac{\hbar\omega_{ho}}{k_B}(N-N_0)^{1/3} \tag{4}$$

From Eqn. (4), it can be seen that if the number of thermal atoms is constant, the temperature of the background gas in the trap depends on the frequency of the trap ($\omega_h$) and the number of condensed atoms ($N - N_0$). Therefore, Eqns. (1) and (4) both indicate that we need to reduce the frequency of the trap to reduce the temperature of the bound atomic gas. Eqn. (4) also reveals that one can continue to reduce the temperature of the system while some of the atoms $N_0$ have already been condensed, signifying that the cooling process can continue even if the space phase density is greater than the critical value, for instance, 1.2 for the harmonic trap.

For a practical experimental system, the trap containing the atomic gas can be either a magnetic trap or an optical trap. We will discuss the optical trap as an example. The ideal simple harmonic trap is shown in Fig. 1.

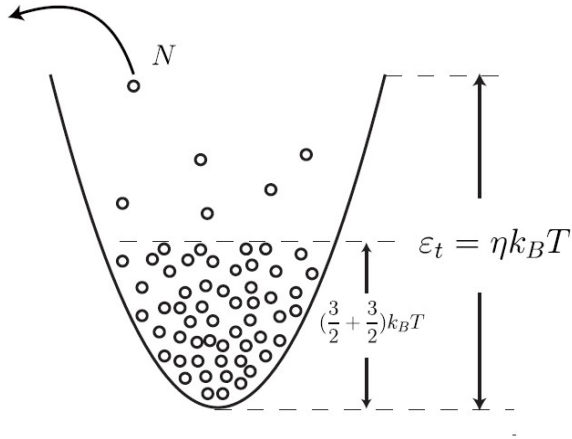

**Fig. 1** Schematic of the harmonic trap with trapped atoms. Evaporative cooling continuously reduces the optical potential barrier and removes atoms with large kinetic energy from the trap.

A common way to further cool the atomic gas after being loaded into the optical trap is forced evaporation cooling. In this process, atoms with high kinetic energy fly out of the trap to remove energy and reduce the temperature of the atomic gas in the trap. If we assume that there are N atoms in the trap, then the average total energy (kinetic energy + potential energy) in the trap is $E = \frac{3Nk_BT}{2} + \frac{3Nk_BT}{2} = 3Nk_BT$, The evaporative cooling of an optical trap is controlled by a time sequence, which causes the atoms with kinetic energy higher than the corresponding energy of the optical barrier (also known as truncated potential energy) to evaporate rapidly. These atoms move out of the trap quickly (see Fig. 1), and the ratio of the energy carried out of the trap to the average kinetic energy inside the trap is the truncation coefficient: $\eta = \frac{\varepsilon_t}{k_bT}$. In the process of evaporative cooling, if dN atoms are evaporated and energy is removed as dNε_t = dNηk_bT, the total energy of the atoms left in the trap is decreased by dE:

$$dE = dN(\eta - 3)k_bT \tag{5}$$

With elastic collision, the temperature T of an atomic sample in the trap decreases by the quantity of dT and becomes T' (see Fig. 1). The following relations can be obtained through energy conservation [62]:

$$3(N - dN)k_BT - dE = 3(N - dN)k_b(T - dT) \tag{6}$$

Eqns. (5) and (6) demonstrate that the relative rate of change between the average temperature T and the number of atoms N of the gas remaining in the trap is constant (i.e., the scaling law) [63–64]:

$$\alpha = \frac{d(\ln(T))}{d(\ln(N))} = \frac{\dot{T}/T}{\dot{N}/N} \tag{7}$$

where $\alpha$ is called the efficiency of evaporative cooling, which is known by theory [64] as α = η/3 - 1. Because $\eta$ remains unchanged during evaporative cooling, $\alpha$ also remains unchanged. The ratio of the temperature T(t) in the trap to the initial temperature T(0) and the ratio of the number of atoms N(t) in the trap to the initial number N(0) after evaporative cooling for the period of time t are as follows:

$$\frac{T(t)}{T(0)} = \left(\frac{N(t)}{N(0)}\right)^\alpha \tag{8}$$

It can be seen from the above that forced evaporative cooling can reduce temperatures by lowering the optical potential and driving away high-energy atoms. In order to continue the evaporative cooling, it is necessary to increase the collision rate of atoms, $\Upsilon_c = n\sigma v$, in the evaporative cooling process (where n is the gas density, σ is the collision cross section, and v is the average relative velocity of atoms in the trap). As time progresses, the runaway evaporative cooling relationship should be satisfied as

$$\frac{1}{\gamma_c}\frac{d\gamma_c}{dt} = -\frac{1}{\tau_{ev}}(1-\alpha) - \frac{1}{\tau_{bg}} > 0 \tag{9}$$

where $\tau_{ev}$ is the evaporative cooling relaxation time and $\tau_{bg}$ is the background gas collision relaxation time. For the experiment, the time sequence design needs to increase the atomic density n at a faster rate while decreasing the temperature T of atoms in the optical trap and the collision velocity v of the atoms. The final cooling effect will lead to the continuous increase in the phase space density of the system and ultimately meet the following requirements:

$$\rho_{ps} = n\lambda^3 = \left(\frac{\hbar\bar{\omega}}{k_B T_c}\right)^3 = \zeta(3) \tag{10}$$

which is the same requirement is as for Eqn. (2).

However, on the ground, the optical trap is not symmetrical. Under the influence of gravity, a symmetrical trap will be transformed into the trap potential shown in Fig. 27 (b) (blue solid line), and a gap will be created in the gravitational direction of the trap potential, from which the atoms will leak out (Fig. 27 (c)). The above analysis tells us that the runaway evaporative cooling requirement Eqn. (9) needs to be satisfied in order to effectively reduce the temperature of the atomic gases, i.e., a high density ($n$) should be maintained during the cooling process. Therefore, in practical experiments, we must make the height of the trap potential barrier higher than the critical value to prevent atoms from escaping from the gap generated by gravity and reducing the density $n$ (detail analysis see section 5.2). If there are $10^5$ Rb atoms, the frequency of the trap used on the ground would be approximately 100 Hz (to prevent atoms from leaking from the trap). From Eqn. (2), we can calculate that the corresponding $T_c$ is around 100 nK, which is consistent with the phase transition temperature in our ordinal laboratory.

According to Eqns. (4) or (2), in order to further reduce the temperature, one must decrease the frequency of the trap $\omega_{ho}$ while maintaining the uncondensed atomic number $N-N_0$. Microgravity conditions provide us with an ideal environment. By eliminating the influence of gravity, the gap in the potential well can be removed. In principle, the optical trap frequency can be reduced to $\omega_{ho}$ = 0.001 Hz, with the corresponding critical temperature of $T_c$ = 1 pK. Therefore, under microgravity conditions, it is possible to obtain ultracold atomic gases of picokelvin magnitude with different traps.

However, if the trap frequency is reduced too much the evaporative cooling process will stop, the atomic collision frequency will decrease rapidly, and the equilibrium time for cooling would need to be greatly increased. For instance, if the trap frequency $\omega_{ho}$ = 0.001 Hz, it would take a collision time over 1,000 s to obtain picokelvin temperatures. Therefore, a novel fast cooling method is needed for these experiments that can also reduce the temperature to picokelvin level.

There are two ways to realize fast cooling after evaporative cooling that could enable atoms to reach picokelvin temperatures: controllable decompression cooling and delta-kick cooling (DKC). The principle of controllable decompression cooling is that the atoms lose energy during adiabatic expansion when the system is decompressed. Decompression cooling in an optical trap is realized by controlling two parameters, laser power and laser beam diameter, which can effectively reduce the potential energy of the atoms during the transformation from one optical trap to another. By controlling the depth and size of optical trap, the atoms with large kinetic energy can be driven out of the trap to rapidly cool the atomic system. Decompression cooling is commonly conducted after evaporative cooling in an all-optical trap, it is also called two stage cooling (TSC). The principle of DKC is to release the atoms in the trap so that the kinetic energy distribution of the atoms can be transformed into the spatial position distribution. After applying an additional pulsed potential well, the atoms in different positions can receive separate forces so that the speed of the atoms in various positions can be reduced to zero. This type of deep cooling procedure could further reduce the temperature of the entire system.

In the process of evaporative cooling, the decrease in the atomic velocity is accompanied by an increase in the phase space density. This means that atoms gather not only in momentum space, but also in position space, and the entropy of the system decreases simultaneously. However, DKC and decompression cooling can reduce the kinetic energy of the atom and concentrate it in momentum space, but the atom expands in position space. Therefore, the phase space density of the system does not increase, and the entropy does not change significantly.

Presently, there are two kinds of traps for deep cooling under the condition of microgravity: an atom chip with a magnetic trap plus microwaves and an all-optical trap with crossed laser beams. The corresponding cooling methods are also different. DKC is primarily used for atom chips, while TSC is mostly used for all-optical traps. Because microgravity experiments need to be carried out on the space station or a rocket, comparisons on the ground are more complex. Several groups have carried out relevant experiments on the ground or in space. We will elaborate on their basic principles and experimental progress in the following sections.

## 3. Ground-based experiments for accessing the picokelvin temperature regime

### 3.1 Achieving 450 pK using Na in a magnetic trap [40]

Section 2 mentions that due to the influence of the gravity of the Earth, the trap frequency of the loaded atoms cannot be too low, so the ultimate cooling temperature limit is about 100 nK. In order to further reduce the temperature, gravity must be counteracted on the ground so that the frequency of the trap can be reduced. In 2002, Ketterle's team conducted their first experiment, reducing the temperature of a confined atomic Na gas to below 500 pK. The key step to achieve temperature of 500pK is that they implemented two stages of decompression cooling after evaporative cooling phase.

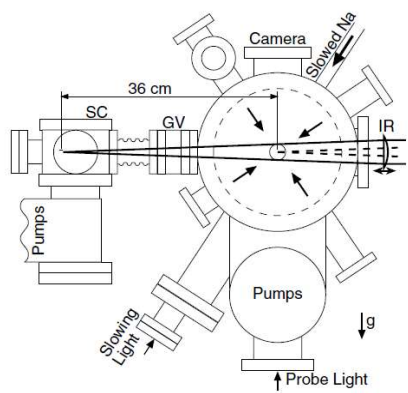 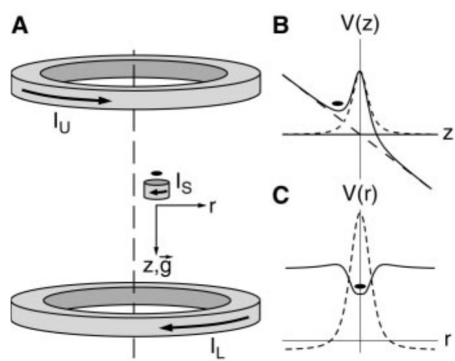

**Fig. 2** Schematic of the apparatus, side view. The science chamber (labeled SC) is isolated by a gate valve (GV) from the trapping chamber[65].

**Fig. 3** Schematic of gravitomagnetic trap. **(A)** Bose-Einstein condensates were levitated against gravity, and the 25-turn coil mounted inside the ultrahigh vacuum chamber running current $I_s$. **(B)** Magnetic potential due to $I_s$,

gravitation potential (short dashed line), gravitation potential (long dashed line), and joint vertical potential of the gravitomagnetic trap (solid line). **(C)** A radially repulsive magnetic potential was generated by running $I_s$ alone (dashed line); however, applying a slight antibias field with $I_u$ modified the radial energy profile and created a magnetic field minimum at $r$ = 0 (solid line) [40].

They trapped sodium atoms using Ioffe-Pritchard magnetic traps and then used radio frequency evaporative cooling to obtain cooler atomic gases. The atomic gases were transported to a separate chamber—the scientific chamber—using optical tweezers [65], a manipulating system for cold atoms using light forces. A coil set with a balanced-gravity functional magnetic field, as shown in Fig. 3A, was installed in the science chamber. The small coils generated a vertical bias field $B_z$ and provided a vertical magnetic field gradient $B'_z$ = 8 G/cm to balance the force of gravity by producing an upward force, suspending the atom 0.5 cm above the small coil. The small coil also produced magnetic field gradients in the radial direction (x, y direction). The final result was a simple harmonic potential away from the center position (Fig. 3B). Two additional coils above and below the small coil generated an additional controlled magnetic field, $B_z$, and a magnetic field gradient, $B'_z$, with currents $I_U$ and $I_L$, respectively; they also produced magnetic field gradients in the radial (x, y) direction. The three-dimensional harmonic magnetic trap with balanced gravity could be generated by adjusting the currents, $I_s$, $I_U$, and $I_L$.

The gravitomagnetic trap was switched on to load the atoms into the chamber with the optical tweezers. The trap was spherically symmetric, with the frequency of $\omega x \approx \omega y \approx \omega z \approx 2\pi \times 8$ Hz. When $5 \times 10^5$, sodium atoms were loaded into the gravitomagnetic trap, a 5 s delay was required to dampen the excited atoms. The critical temperature of the BEC was then $T_c$ = 30 nK. In the experiments, the temperature $T$ of atomic gases were always maintained $0.5 < T / T_c < 1$.

Further cooling consisted of two decompression processes of 5 s each. In the first stage, the current of the small coil $I_s$ was reduced to 1/10, and the currents of the other two large coils, $I_U$ and $I_L$, were increased simultaneously so that the longitudinal frequency of the potential well was reduced to $\omega_z = 2\pi \times (1.81 \pm 0.05)$ Hz. Thus, the primary gradient effect of the magnetic well was transferred from the small coil to the two additional large coils. In the second stage, the radial frequency of the trap was reduced by increasing $I_L$ and decreasing $I_U$ by the same amount. The three currents were set so that the final gravitomagnetic trap frequency was $\bar{\omega} = 2\pi \times (1.12 \pm 0.08)\ Hz$ and the axial bias magnetic field was $B_z$ = 17 G. Following this decompression process, there were $2 \times 10^5$ condensed sodium atoms with the critical temperature of $T_c$ =3 nK. According to Eqns. (4) or (2), if the number of atoms is further reduced, the temperature would also decrease. In the experiment, the elastic collision rate (between thermal atoms and the condensates) decreased from 0.25 Hz to 0.01 Hz through 10 s of microwave radio-frequency evaporation, with a substantial decrease in the collision rate for thermal equilibrium and the rate of evaporative cooling. Nevertheless, when the number of atoms dropped to 30,000, the temperature of the atomic gas was below 1 nK (Fig. 4A), if the number of atoms was further reduced to 16,000, the temperature of the atomic gas would be reduced to 780 ± 50 pK (Fig. 4B). The lowest temperature obtained was 450 ± 80 pK (Fig. 4C) with 2,500 atoms remaining. The relationship between the temperature and the number of atoms in the trap is shown in Fig. 5.

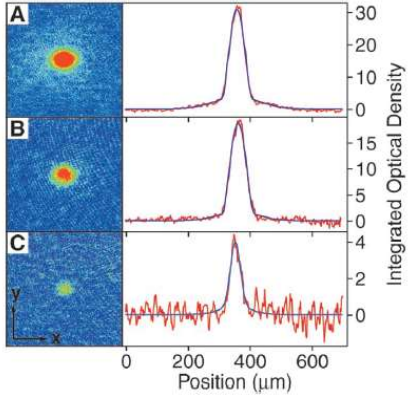

**Fig. 4** Picokelvin temperature thermometry. Partially condensed atomic vapors confined in the gravitomagnetic trap with **(A)** 28,000, **(B)** 16,000, and **(C)** 2,500 atoms [40].

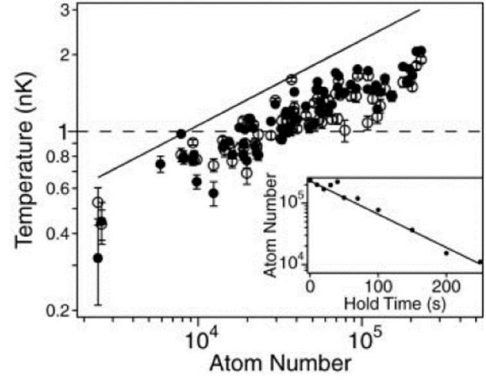

**Fig. 5** Bose-Einstein condensates at picokelvin temperatures. The temperature of more than 60 partially condensed atomic vapors is plotted versus total number of condensed and non-condensed atoms. A solid line at the Bose-Einstein condensation phase transition temperature and a dashed line at 1 nK are provided as guides [40].

When the temperature of an atomic gas is lower than 1 nK, it is difficult to measure the temperature by the time-of-flight (TOF) technique. While detecting the atomic vapor temperature with TOF on the ground, atoms fall out of the CCD detection area due to gravity after 20 ms. From Table 1 in Section 5.2, it can be seen that the size of the atom cloud only changed by 16 μm in 10 ms when the temperature was less than 10 nK. Therefore, it is difficult to obtain an image with CCD when the temperature is lower than this.

For this reason, they used Eqn. (3), which could be used when the number of atoms $N_0$ of the condensate and the background thermal atoms $N_{th}$ were measured, to determine the temperature. In the experiment, the critical temperature for condensation $T_c$ was measured initially. The atomic velocity distribution transferred in space is shown in Fig. 4 A, B, and C. The temperature of the atomic vapors was extracted by fitting the bimodal distribution (Fig. 4) using Eqn. (11):

$$n(x) = N_0 \Psi_0^2 + \frac{N_{th}}{\sqrt{\pi} w_{th}} e^{-\frac{x^2}{w_{th}^2}} \qquad (11),$$

where $\Psi_0^2$ is a bell-shaped function, whose width $w_0$ describes the peak value of the condensate $\Psi_0^2 = (\frac{15}{16}) w_0^{-1} \max(1 - \frac{x^2}{w_0^{-1}}, 0)^2$ of a Thomas-Fermi gas $\Psi_0^2 = w_0^{-1} \pi^{-1/2} e(-\frac{x^2}{w_0^2})$ for an ideal gas. Fig. 4 demonstrates that there is a two-mode density distribution in the condensate. The temperature information could be obtained by fitting the experimental data with Eqn. (11), which is consistent with the temperature determined by Eqn. (3). Finally, the lowest temperature of 450 pK was achieved.

**3.2 Transverse cooling to 50 pK by delta-kick cooling [45]**

In addition to obtaining ultralow temperature atomic samples in ground-based experiments by forced evaporative cooling while balancing gravity with magnetic forces, a separate method for cooling is the DKC method. In 2015, the Kasevich team achieved a transverse temperature of 50 pK for falling atoms using this method. DKC is a new cooling method for sub-photon-recoil temperatures proposed by Hubert Ammann and Nelson Christensen in 1997 [44]. The basic idea of the method is to trap atoms using an optical trap, then suddenly switch off the trap for time of *t*, during which, the atoms will freely diffuse. Subsequently, the optical trap is turned on for a short time of $\delta t$ so that the diffused atomic gas is cooled by a force opposite to the direction of diffusion. Compared with evaporative cooling, DKC cooling has the advantages of short cooling time and low atomic number loss. We can use a simple theory to describe its principle.

Assuming that an atom of mass *m* is bound in the trap potential *U(x)*, the Hamiltonian can be written as [44]

$$H_0 = p^2 / 2m + U(x) \qquad (12).$$

This represents the atomic state at *t* < 0. At *t* = 0, the trap potential *U(x)* is turned off, and then at time $t_0$, the trap potential *U(x)* is turned on for a short interval of $\delta t$. The process can be described by $U(x)\exp[-(t-t_0)^2/2\tau_p^2]$. If the time interval $\delta t$ is short enough, then the Gaussian pulse is equivalent to a delta function $\delta t$. Therefore, the Hamiltonian of the system for *t* > 0 can be expressed as

$$H_k = \frac{p^2}{2m} + V(x)\delta(t - t_0), \qquad (13)$$

where $V(x) = \sqrt{2\pi}\tau_p U(x)$.

By Eqn. (13), it is evident that the strength of the impact can be adjusted by changing the pulse width $\tau_p$. Assume that atoms begin to accumulate at the minimum of the trap potential *U(x)*. If the trap is turned off, after $t_0$ s, atoms of different momenta will reach separate positions in space so that the momentum *p* becomes a linear function of the position *x*: $P(x) = mx/t_0$. An applied pulsed optical trap can be approximated using a simple harmonic well: $U(x) \approx m\omega^2 x^2/2$. This pulsed trap will change the momentum of the atom as: $\Delta p \propto dU/dx \propto x \propto p$. Consequently, during the time interval $\delta t$, the force acting on the atom at *x* is $F(x) = -\frac{dU}{dx} = -m\omega^2 x$, meaning that an atom at position *x* will have an acceleration of $-\omega^2 x$. It is assumed that the atoms do not move significantly during the time interval $\delta t$, so the change in the atomic momentum is:

$$F(x)\delta t = -m\omega^2 x\, \delta t = p(x, t_0 + \delta t) - p(x, t_0) \qquad (14),$$

where $p(x, t_0 + \delta t)$ is the final momentum of the atom after the pulsed trap is applied, and $p(x, t_0)$ is the initial momentum of the atom before the pulsed trap is applied, from Eqn. (14), if the following conditions are met:

$$\omega^2 t_0 \delta t = 1 \qquad (15).$$

The final momentum of the atoms can be represented by $p(x, t_0 + \delta t) = mx\left(\frac{1}{t_0} - \omega^2 \delta t\right) = 0$.

Thus, the kinetic energy of the atom will be reduced to zero. The energy of atoms decreases to zero when only those atoms at the origin meet the conditions of Eqn. (15). By the above analysis, the DKC process clearly cools atoms by converging their momenta to zero. This effect is similar to an optical lens; therefore, the DKC process is also referred to as the "lens" effect in some literature. For radial cooling, DKC is also called "lensing collimation" [67–74]. There are three kinds of trap potentials that can generate lens pulses: magnetic [69, 75–76], electrostatic [77], and optical dipole traps (ODT) [45, 67]. The ODT was used in the experiments of the Kasevich team [45].

In the Kasevich experiment, the atoms at the bottom of the trap distributed in a certain range around the origin; therefore, the conditions of Eqn. (15) could not be fully satisfied. Additionally, even if atoms are concentrated at the origin of the trap potential, it is difficult to obtain an ideal harmonic trap because there will always be additional harmonic components. One can obtain an intuitive picture of the momentum or velocity distribution of atoms in phase space. Ammann in reference [44] calculated the variation in the momentum $\sqrt{\langle p^2 \rangle}$ and position $\sqrt{\langle x^2 \rangle}$ distributions of atoms in the DKC process (see Fig. 6). In phase space, the atoms around the origin are initially distributed spherically about the origin. After $t_0$ s of free evolution following shutdown of the trap, the atomic spatial distribution becomes wider and the distribution in phase space becomes obliquely elliptical. During the time interval $\delta t$ of the pulse, the momentum distribution reduces, while the distribution in phase space becomes a positive ellipse, which indicates that the momentum or velocity distribution is compressed, and the position distribution remains unchanged. Therefore, the phase space density (PSD) does not change during the DKC process. In phase space, the longer the diffusion time $t_0$, the wider the position is spread, the narrower the momentum distribution is compressed, and finally, the more the temperature is decreased (Fig. 6).

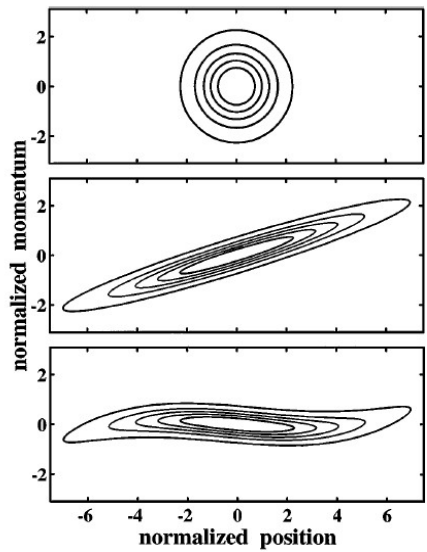

**Fig. 6** Evolution of the classical phase space density: initial (top), after free expansion (middle), and after the kick (bottom) [44].

Fig. 7(a) is a diagram of an experimental setup consisting of an atom interferometer with a 10 m vertical vacuum tube. The atomic source was generated from a time-orbiting potential (TOP) trap, a vertically oriented quadrupole trap, and a horizontal top coil pair. A group of $10^5$ Rb atoms were evaporatively cooled by TOP to achieve an effective temperature of 1.6 ± 0.1 nK. Using a chirped optical lattice [79, 80] (blue detuning optical lattice with the upward arrow in Fig. 7(a)) to project cold atoms upward, and a 3 W laser is mirrored from bottom to the top and interacts with the cold atomic cloud (for clarity, the angle between the reflected laser beams is 1 mrad, and the beam angle is magnified in Fig. 7(b)). The reflected laser acts as a dipole lens, and the dipole lens potential [68] is produced by the transverse intensity distribution of the Gaussian beam, providing two-dimensional cooling. The laser has a waist of 3.4 mm, and the frequency is detuned to 1.0 THz with respect to the $^{87}$Rb $D_2$ line. The atoms are launched upwards into the 10 m vacuum tube. After 2.8 s, the atoms fall back to the detection zone where they are imaged by two CCD cameras with a vertical fluorescent beam (y-axis camera imaging the x-z plane, x-axis camera imaging the y-z plane). Fig. 7(c) is the fluorescence image of the cloud after launching the atoms upward. Fig. 7(d) shows the spatial distribution of the atoms refocused with a dipole lens from Fig. 7(c). By analyzing the interference pattern, the effective temperature was found to be 1.6 nK (Fig. 7(c)).

After the optical trap is switched off, the atom is launched forward. In addition to free diffusion, the atoms will fall freely due to the influence of gravity. The falling distance is approximately 5 m in 1 s. Therefore, in order to position the atoms in the region of optical trap, the diffusion time $t_0$ can only be about 10 ms. Therefore, the efficiency of DKC cannot be very high, and significantly low temperatures cannot be reached. In 2015, the Kasevich team achieved a long-term drop process to obtain a long diffusion time $t_0$ for free-fall atoms, during which radial DKC was performed such that the atomic free-diffusion time $t_0$ was greater than 1 s, achieving an ultralow transverse temperature.

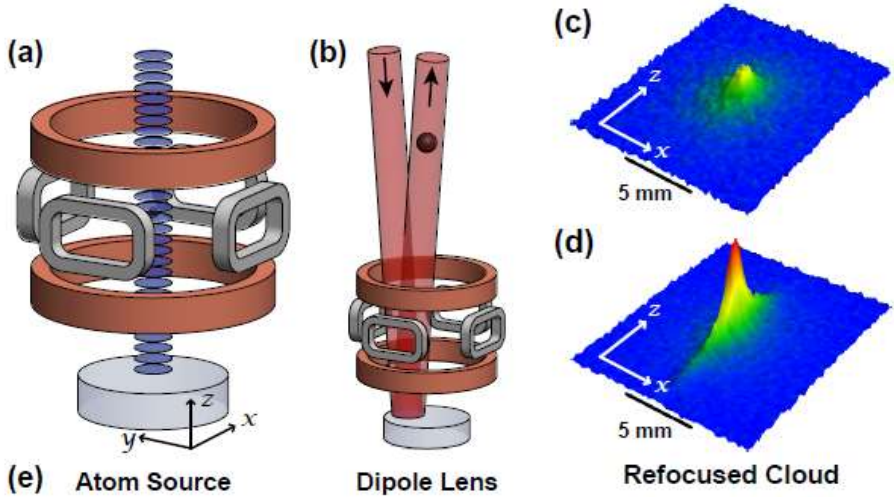

**Fig. 7 (a)** Schematic of the apparatus (including vertically oriented quadrupole trap, horizontal TOP coil pairs, and blue-detuned launching lattice), **(b)** A 3 W laser, $1/e^2$ radial waist σ = 3.4 mm, 1.0 THz red detuned from the $^{87}$Rb $D_2$ line, acts on the atom cloud as a dipole lens (the ~1 mrad beam angle is exaggerated for clarity), **(c)** Fluorescence image of a 1.6 nK cloud after 2.8 s of free fall, **(d)** The distribution in **(c)** refocused using the dipole lens [45].

At the beginning of the experiment, the size of the atomic cloud is $\Delta x_o$, and the velocity distribution is $\Delta v_o$. The atom is freely diffused in the radial direction during the launching upward process. After the diffusion time $t_0$ (also referred as the object time), an optical trap pulse was applied (herein referred to as: lens is applied and the lens time depicted as $\delta t$). At this point, the size of the atomic cloud becomes $\Delta x_l$, the velocity distribution becomes $\Delta v_1$, and the corresponding temperature ratio is $\eta \equiv (\Delta v_1/\Delta v_0)^2$. For an ideal simple harmonic trap, the minimum velocity distribution can be obtained as $\Delta v_1$ (or collimation condition), but the corresponding temperature has a lower limit than $\eta_c$, which is $\eta_c \equiv (\Delta x_0/\Delta x_l)^2$. From this equation, we know that the atomic cloud needs to have a larger diffusion scale and a longer diffusion time to acquire a lower temperature: $\Delta x_l \approx \Delta v_0 t_0 \gg \Delta x_0$. After the diffusion, the optical potential well is turned on, and the force acting on an atom is $F(x) = -m\omega^2 x$, and the velocity changes to $\delta v(x) = -\omega^3 \delta t\, x$. In this way, if the atom source is a point atom, when the diffusion time of the atom is $t_0 = 1/\omega^2 \delta t$ (for a 2D case, which is also called the collimation condition), the ideal collimation of the atomic cloud will be obtained; that is, the transverse effective temperature of the atomic group tends to absolute zero. Because the effect of the trap potential on the transverse velocity of the atom is similar to that of the collimation of the optical lens, the optical trap potential effect is called the lens effect in [45].

As mentioned above, when the temperature of the atomic cloud is less than 1 nK, it takes a long time to observe the scale change of the atomic cloud, which takes about 10 s. To overcome this difficulty, the turn-on time of the optical trap was extended so that the lens time was greater than $\delta t = \frac{1}{t_0}\omega^2$ and the atomic cloud would be re-focused. Like geometric optics, the smallest size of imaging is the focus point of the atoms after cooling (or collimation). This method has been used for temperature measurements of electron beams [81].

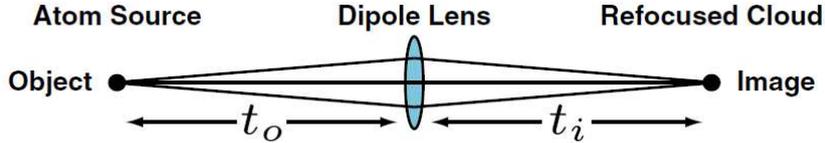

**Fig. 8** Optical analogy showing the object, lens, and image, with object distance $t_0$ and image distance $t_i$ [45].

Fig. 8 is a diagram corresponding to this measurement. Atoms fall from the starting point (called the object). After the diffusion time $t_0$, the encounter of the lens (optical trap potential) within the time interval $\delta t$ (equivalent to the thickness of the lens) begins to converge. After the time $t_i$, it converges into an atomic cloud on the image side. In fact, the minimum size of the image $(\Delta x_i)_{min}$ corresponds to the minimum velocity distribution $\Delta v_l$:

$$(\Delta v_l)^2_{bound} \equiv \frac{(\Delta x_i)^2_{min}}{t_i^2} = \Delta v_l^2 + \delta A \geq \Delta v_l^2 \tag{16}$$

Here $t_i$ is called the image time and $\delta A$ represents the aberrations of the lens. Because $\delta A$ is positive, $(\Delta v_l)^2_{bound}$ is the upper bound of the collimation temperature. In the experiment, $t_0$ = 1.1 s and $t_i$ = 1.8 s, and with different lens durations $\delta t$, one can get different sizes of atom images.

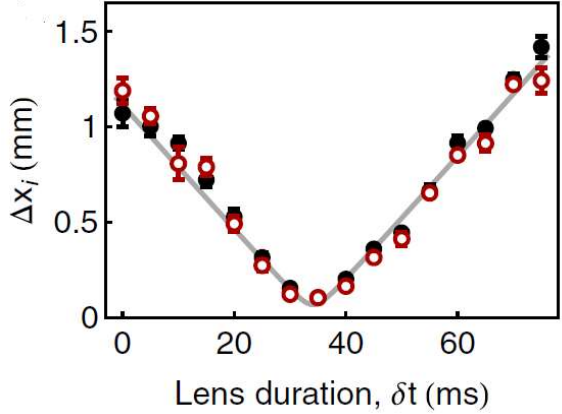

**Fig. 9** Filled black (open red) points denote measured rms cloud widths on the x-axis (y-axis) camera. Each point is the weighted mean of Gaussian fits to six experimental shots. The dashed gray curve is a simultaneous fit to the measurements from both cameras and reports a minimum size of 70 µm at a lens duration of 34 ms [45].

Fig. 9 shows the change in the lens duration $\delta t$ corresponding to the size of the atomic cloud on the image side. When $\delta t$ = 35 ms the smallest image size is obtained: $(\Delta x_i)_{min} = 70~\mu m$. The upper limit of the transverse collimation temperature can be estimated by Eqn. (16) [82]. For the x-axis, $(\Delta v_i)_{bound} = (\Delta x_i)_{min}/t_i = 65 \pm 20 \mu m/s$, and for y-axis, $(\Delta v_i)_{bound} = 75 \pm 25 \mu m/s$. The effective temperature should be $T_{bound} \equiv m(\Delta v_l)^2{}_{bound} = 40^{+40}_{-20}~pK, 50^{+5}_{-3}~pK$, corresponding to x-axis and y-axis, respectively. The effective temperature is the transverse temperature in the x and y directions, respectively, and the temperature in the gravity direction (z direction) is still approximately 1 nK. Fig. 7(c) is a TOF image of a 1.6 nK atomic cloud at 2.8 s after it was launched upward, and Fig. 7 (d) is a TOF image of the atomic cloud with a lens process during a 2.8 s the upward-launching phase. The temperature in the x direction is reduced (estimated to be around 40 pK), and there is no significant change in the z direction. The overall effect is that the peak of the atomic distribution becomes higher, which help to improves the SNR for the next step of the experiment.

## 4. Accessing the picokelvin regime using gravity-free techniques

The lowest temperature achieved by a ground-based experiment was 500 pK [40]. Further progress would require gravity-free or microgravity methods using a satellite, rocket, or orbing space station. The trap (including the optical trap and magnetic trap) depth can be further reduced to less than 1 Hz, with the goal of obtaining a temperature below 500 pK. Second, with microgravity, the atom is almost stationary, and the preparation and detection times can be increased. For DKC, it is easy to obtain a longer release time or lens time under microgravity conditions and increase the measurement or image time. There are currently two schemes for achieving lower temperatures. The first is an atom chip technique, using DKC or adiabatic release, currently employed by German and American teams. The second is all-optical trap technique, using a two-stage cross-beam cooling method, referred to as TSC or TSCBC, primarily used by French and Chinese teams.

### 4.1 Experiments for QUANTUS and MAIUS-1

Atom-chip cooling employs a highly miniaturized magnetic trap formed by conductors on the surface of a microchip. The chip typically traps $10^4$ to $10^7$ atoms. The experimental study of atom-chip cooling of atoms under microgravity conditions was first carried out by the Rasel group at the University of Hannover. With the support of the German Space Agency (DLR), they conducted a series of experiments on the QUANTUS project. In 2010, they implemented BEC on the drop tower at the Center of Applied Space Technology and Microgravity (ZARM) in Bremen [47]. The temperature of the atomic gas was around 9 nK. In 2013, they successfully used the DKC process on the drop tower, obtained BEC, and reduced the temperature of the atomic gas to about 1 nK [49]. They also conducted an atomic interference experiment based on BEC. In 2017, they employed DKC on a sounding rocket and obtained BEC, reducing the temperature of the atomic gas to about 1 nK with an extended experiment time [52]. In 2018, NASA's Jet Propulsion Laboratory installed the Cold Atom Lab (CAL) on the ISS, and performed $^{87}$Rb, $^{39}$K, and $^{41}$K gas cooling experiments using an atom chip to obtain an extended experimental time [54–56]. The goal of the above experiments was to break the limit of 500 pK minimum temperature on ground. The experiment is still being further improved with the goal of entering the picokelvin temperature regime in the microgravity environment. The specific experimental progress of the above two experimental teams is described below. Both the German and American teams use atom chips, a solution that uses a magnetostatic trap plus microwave for cooling.

Compared with traditional BEC magnetostatic trap experiments, the atom chip technique has several advantages [46]. First, approximately 1 kW of power is required to achieve all the magnetic field configurations required to generate BEC with coils in conventional setup, while an atom chip only needs 10 W. However, the coils used for the external bias field and the outer quadrupole coils of magneto-optical trap (MOT) of atom chip consume approximately 350 W of power. Second, the limit in the magnetic trap generated by the atom chip is an order of magnitude stronger than that of a typical magnetic trap in conventional system, so the rethermalization speed is faster, and the evaporation time of the magnetic trap realized by the atom chip is shorter. Third, experiments of ultracold atoms in space must consider the miniaturization of the system. Therefore, atom chips are one of the best options [83–85].

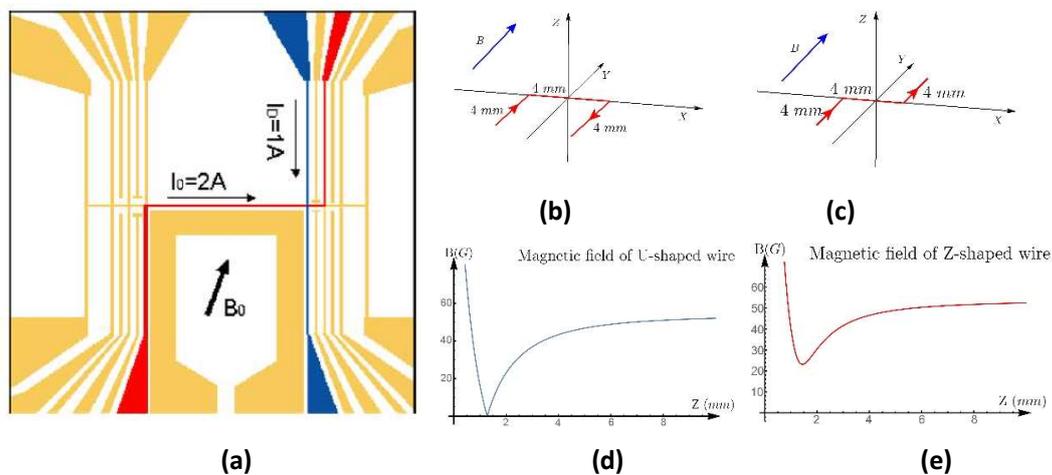

**Fig. 10 (a)** Layout of the atom chip, the bottom yellow wide wire in the center is the U-Shaped wire, and the red wire is the z-shaped wire [46]. **(b)** U-shaped wire plus a magnetic field above the surface of the chip. **(c)** Z-shaped wire plus a magnetic field above the surface of the chip. **(d)** The magnetic field in z direction for U-shaped trap, the same as in x, y directions. **(e)** The magnetic field in z direction for Z-shaped trap, the same as in x, y directions.

Atom chips are microfabricated, integrated devices in which magnetic, electric, and optical fields can confine, control, and manipulate cold atoms. A magnetic trap is a commonly used technique to trap atoms, and most magnetic traps are made of large-scale electric-current-driven coils. The micro magnetic trap of atom chip uses the etching technology of microelectronics, which etches the metal wire on the chip of several square centimeters (Fig. 10 (a)) and combines the thin wire on the chip with the external magnetic coils above the chip to form the magnetic trap. The outstanding advantage of this kind of magnetic trap is that it can produce a considerable magnetic field gradient with a low current and form different configurations of trap potential wells. It can produce magnetic field gradient of 500 G/cm when the current is only 1 A and the magnetic field is only 10 G. The most prominent feature of atom chip is that it can easily generate a larger magnetic field gradient comparing with standard magnetic coils.

There are two wire configurations on atom chips to produce different magnetic traps: U-shaped and Z-shaped. U-shaped wire plus a pair of coils, which produce parallel magnetic field, can form a quadrupole trap (Fig. 10 (b)) for the operation of an on-chip MOT with the help of laser beams and produce the magnetic field shown in Fig. 10 (d). Z-shaped wire plus a pair of coils can form an Ioffe trap (Fig. 10 (c)) and produce the magnetic field shown in Fig. 10 (e). In comparing the two wire traps, it can be seen from Fig. 10 (e) that there is no zero point of the magnetic field for the Z-shaped wire trap. This trap can produce the Ioffe field for the operation of evaporative cooling with the help of an RF field. Because the atom chip can produce a tightly bound magnetic trap near the surface, it provides a fast and efficient way to generate BEC. After obtaining BEC, the chip trapping potential can be significantly reduced and the trapped atoms can be separated from the surface to avoid surface effects, which usually disturb the trapped atoms.

Fig. 10 (a) shows the scheme of the atom chips used by the German team. They selected a magnetic trap on the chip to meet the small size and low power requirements. The chip contained a U-shaped wire that produced a quadrupole field of the MOT and a Z-shaped wire for generating an Ioffe-type magnetic trap. An additional straight wire created a strongly confining 'dimple' in the weak axis of the Ioffe trap and also significantly reduced the trap aspect ratio. The dimple trap provided a trapping frequency of about (1, 3, 3) kHz through the Z-shape and the straight line in the (x, y, z) direction by a current of 2 A and 1 A respectively and allowed evaporative cooling in 1 s.

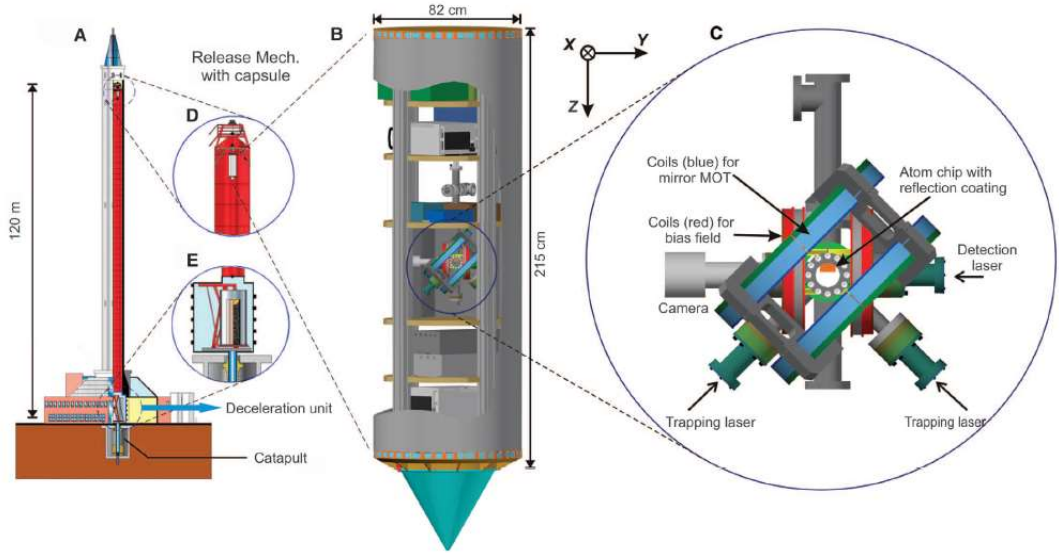

**Fig. 11** Facility of ZARM drop tower in Bremen **(A)** and the capsule **(B)** containing the heart of the BEC experiment **(C)**. The capsule is released from the top of the tower **(D)** and is recaptured after a free fall of 4.7 s through an evacuated stainless steel tube at the bottom of the tower by an 8-m-deep pool of polystyrene balls (E) [47].

The 2010 drop tower experiment used a 146 m high tower with a useful drop time of 4.6 s. The atom chip was mounted in the vacuum chamber of Fig. 11C, which was fixed in the drop capsule (Fig. 11B) [47]. The atom chip was combined with a mirrored MOT, which was loaded with approximately $1.3 \times 10^7$ $^{87}$Rb atoms from the background gas. The MOT required 10 s to load the atoms after which the drop capsule was released. The microgravity environment provided an effective acceleration of $a = 10^{-4}$ $g$. In the first second of free fall, the atoms were fixed in the mirrored MOT while the initial vibration of the drop capsule weakened. The atoms were further cooled with the optical molasses process and transferred to the Ioffe-Pritchard trap on the chip (Fig. 12 A). The atoms were prepared in the hyperfine state with $F = 2$, $m_F = 2$ (where $F$ and $m_F$ are the total angular momentum and the magnetic quantum number of the Zeeman sub-level, respectively). A BEC of about $10^4$ atoms is usually obtained after 1 s of magnetic trap compression and RF-induced evaporation. The lifetime of the condensate in the potential trap is 3 s. The BEC is released by turning off the magnetic field current and then observed using absorption imaging techniques (Fig. 12B). In the experiment, the kinetic energy of the atom was quiet low (about 9 nk). After 1 s of expansion time, the BEC was observed to form a non-localized wave packet extending more than 2 mm; this was the first experiment to obtain BEC using a drop tower. The total time from the capsule release to the final observation was 3 s. Because the evaporative cooling time is limited to 1 s, further cooling to lower temperatures is limited. Nevertheless, the experiment provided strong evidence for the effectiveness of the atom chip in BEC-related technology under microgravity conditions.

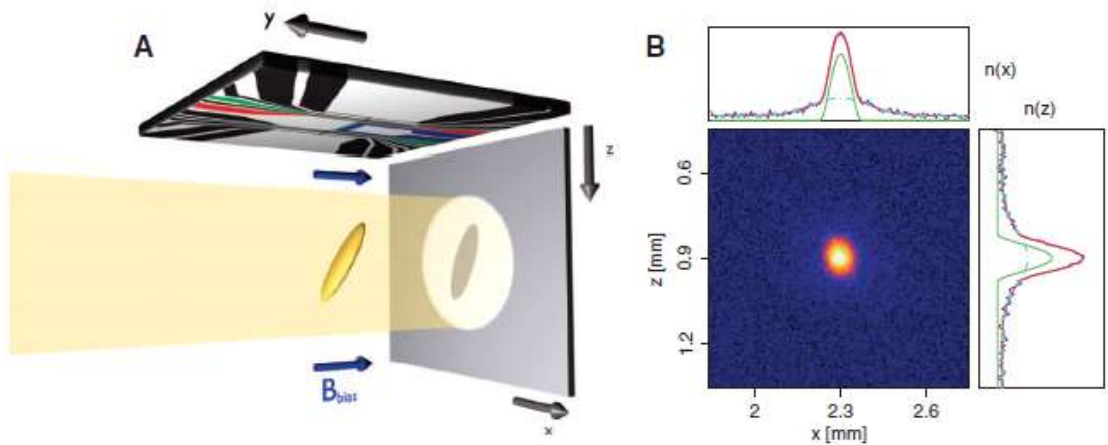

**Fig. 12** Absorption imaging technique **(A)** used for the measurement of the two-dimensional spatial distribution of the BEC in microgravity **(B)**. The resulting two-dimensional density distribution **(B)** corresponds to a BEC created and observed in free fall for an expansion time of 100 ms [47].

As an extension of the Quantas program, the German team conducted a BEC-based interference experiment at the drop tower in the University of Bremen's Center of Applied Space Technology and Microgravity (ZARM) in 2013 [49]. The drop time was increased to 4.7 s, allowing additional time to implement the DKC approach in the cooling process [67–68, 70]. By coupling the Zeeman level to a chirp RF pulse (adiabatic fast passage), the BEC was transferred to the non-magnetic state of $F$ = 2, $m_F$ = 0. This effectively eliminated the harmful effects of residual magnetic fields [87]. Through the above measures, the temperature of the atomic gas was reduced to about 1 nK. An atomic interference experiment was also performed using the ultracold BEC to improve the SNR.

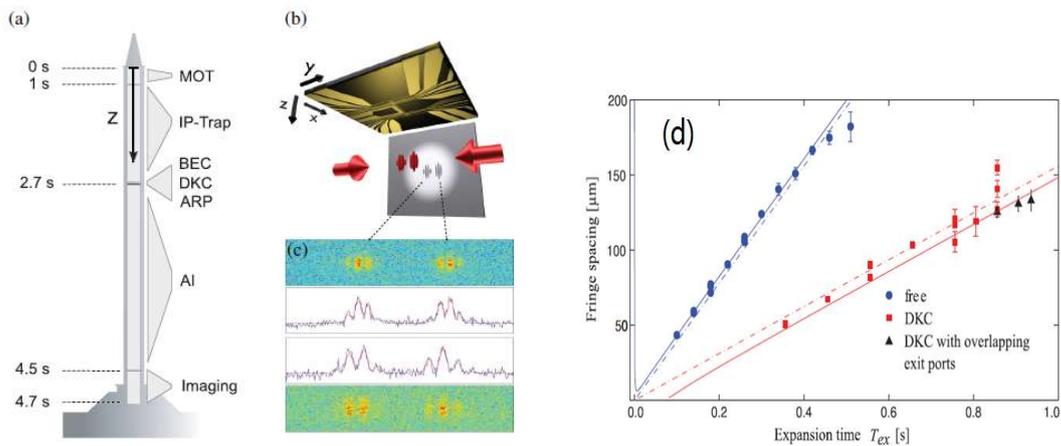

**Fig. 13 (a)** Mach-Zehnder interferometry of a BEC in microgravity as realized in the ZARM drop tower in Bremen **(b)** absorption imaging brings out **(c)** the interference fringes. **(d)** Fringe spacing of two interfering BECs observed at each exit port of an AMZI with (red squares, black triangles, solid red line) and without (blue dots, solid blue line) DKC as a function of the expansion time $T_{ex}$ [49].

Fig. 13 illustrates the 2013 microgravity experiment [49], with Fig. 13 (a) showing the complete time series. The difference from the 2010 experiment was that the DKC process was employed. After evaporative cooling, the trap potential was turned off and the atoms were freely released for 30 ms. The trap potential was then turned on for 2 ms (DKC pulse), with the frequency of the trap of (10, 22, 27) Hz. This atom was further cooled and finally achieved a temperature of 1 nK. In a subsequent interference experiment, an image based on a lower temperature was obtained. The release time of the interference experiment was prolonged, yielding a high SNR (Fig. 14a).

The interference experiment was carried out using the protocol of the Asymmetric Mach-Zehnder interferometer (AMZI) [88–89]. According the AMZI time series in Fig. 14, after the BEC was released for $t_0$ s, the laser was irradiated three times and the single laser pulse absorption imaging detection shown in Fig. 14 (b) was used to detect the atomic interference image. Fig. 14 (c) shows a typical image and corresponding density distribution for two different BEC interferences.

Fig. 13 (d) shows the spatial period of the observed fringe pattern as a function of the extension time *t*. The experimental results (blue circles, red squares, black triangles) are compared with the corresponding theoretical predictions (blue solid lines and red solid lines). The blue line is derived from a model based on the scale method [90–93] and describes the interference pattern of two condensates originally separated by a distance *d*. Their initial shape is derived from a numerical model of the atom chip potential. At large time scales, the observed fringe spacing (blue dot) is identical to the theoretical model [90] and the linear far-field prediction of the double slit (blue dotted line curve). These experiments show that the DKC process under microgravity conditions provides a clear atomic interference fringe, demonstrating that ultralow temperature atoms at the picokelvin level can be achieved with further improved system performance.

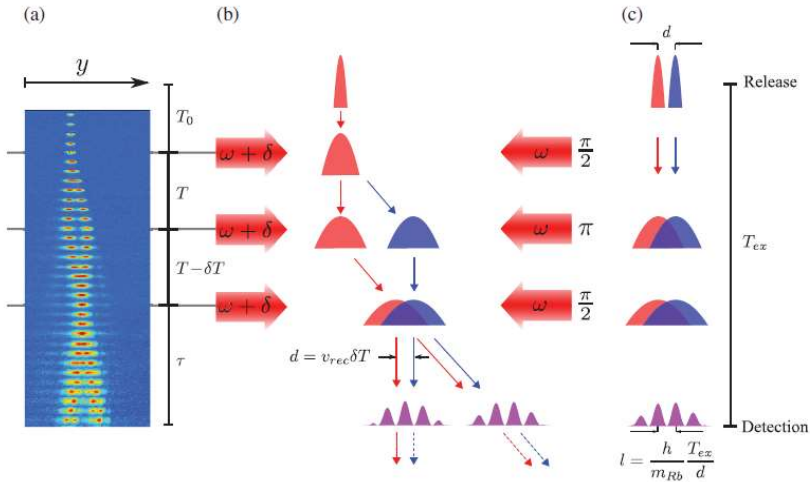

**Fig. 14** Temporal AMZI for a BEC based on Bragg scattering at a light grating: **(a)** experimental images on ground, **(b)** schematic sequence, **(c)** analogy to the Young double-slit experiment [49].

In order to further extend the cooling time, on January 23, 2017, the German team used a sounding rocket to perform a 6-min microgravity experiment in space [52]. As part of this mission, MAIUS-1, they obtained BECs in a microgravity environment to study the phase transition from a thermal ensemble to a BEC. They also studied the collective dynamics of the condensate and matter wave interference. Because the sensitivity of measuring the inertial force with the matter-wave interferometer is proportional to the square of the time it takes for the atom to travel in the interferometer, the free-fall time for the atom in the interferometer under microgravity greatly impacts the measurements. The goal of the experiment was to explore whether Bose-Einstein condensation can be combined with DKC collimation techniques to obtain a slowly expanding ensemble with pico- or femto-kelvin temperatures. This method would allow the atoms to remain in the interferometer for an extended period, improving the precision of the interferometry. This experiment established the possibility of performing quantum gas experiments under microgravity conditions of satellite-borne BECs.

Fig. 15 shows the timetable for the Maius-1 sounding rocket mission. There were three phases: boost (lower left corner), 6 min of microgravity (blue shaded area), and re-entry and landing (lower right corner). The test included 6 min of microgravity, and 110 related atom-optics experiments were performed; the experiments discussed here are marked in red. Above the Karan line —100 km off the ground—, inertial disturbances are reduced to millionths of the gravity on the ground.

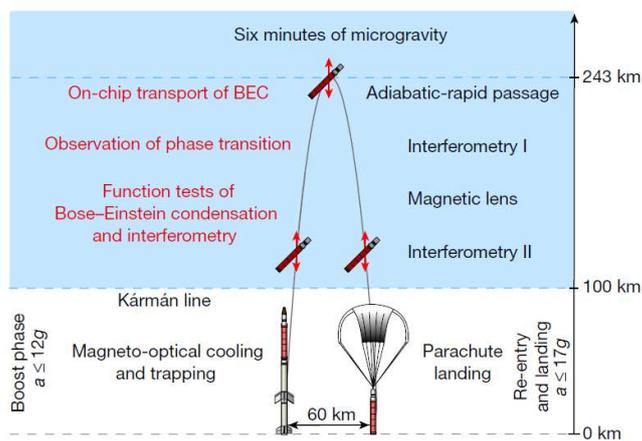

**Fig. 15** Schedule for the MAIUS-1 sounding-rocket mission. Reprinted with permission from [52]. Copyright (2018) by Springer Nature [52].

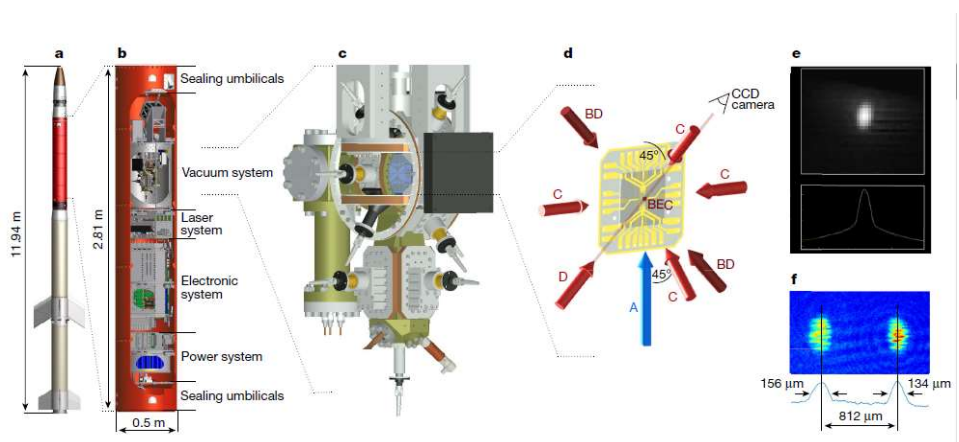

**Fig. 16** Set-up for space-borne Bose–Einstein condensation. The rocket **(a)** carried the payload **(b)**, including the vacuum system **(c)** that houses the atom chip **(d)**, into space. On the atom chip, a magneto-optical trap formed by laser beams **(C)** is first loaded from the cold atomic beam **(A)**. **(e)** Grey-scale absorption image of the spatial density of the BEC in space and its one-dimensional density profile, which were sent to ground control in low resolution. **(f)** Demonstration of Bragg scattering, apparent in the momentum distribution of the BEC [52].

Fig. 16 depicts the launch rocket and the load [94–96]; (a) is the rocket, (b) is the payload, (c) is the vacuum system, (d) is the core part of the experimental device, a multi-layered atom chip [97–100], (e) is an image of the BEC in space, and (f) is a Bragg scattering image of the BEC. On the atom chip a MOT is formed by the laser beam (C) and magnetostatics field from loading atomic beam (A). Subsequently, the magnetic trap on the atom chip performs evaporative cooling to produce the BEC, which is then transported to the region generating interferometry and imaging. Two additional optical lasers (BD) illuminate the BEC to produce Bragg diffraction, and the laser beam (D) illuminates the BEC to produce an absorption image, which is recorded by a charge-coupled device (CCD) camera. Fig. 16 (e) is the visible gray-scale absorption image of the BEC in spatial density (top: white corresponds to the highest density) and its one-dimensional density profile (bottom: integrated from top to bottom of the image) that was sent to the ground control station with low resolution. Fig. 16 (f) is the Bragg scattering, with an observable BEC momentum distribution. The relative time interval of the two BEC peaks is 70 ms, and the gradation shows the spatial density of the cloud (blue for low density; red for high density).

This experiment studied the phase transition of a thermal atom ensemble to a BEC in space for the first time. The temperature of the atoms was cooled by forced RF evaporation, and the BEC was detected in the atom chip magnetic trap. Fig. 17 (a) demonstrates the spatial atomic density of the thermal ensemble and BEC at three different final radio frequencies during forced evaporation (in the final cooling phase). During the phase transition, with the temperature decreasing, the number of atoms in the thermal ensemble (using Gaussian fitting extraction, the red curve in Fig. 17 (a)) is significantly reduced, while the number of atoms in the BEC is significantly increased (parabolic fit, the blue curve in Fig. 17 (a)). Fig. 17 (b) and (c) show a comparison of the formation of the BEC in space and on the ground and also plots the composition of atoms in the BEC relative to the total number of atoms. Comparing the results in Figs. 17 (b) and (c), one can easily find that at the same RF frequency, the observed ratio of thermal atoms and condensed atoms is lower in space than on the ground. This difference may be due to changes in the magnetic field in space relative to that on the ground. Additionally, the number of atoms in the thermal ensemble and BEC in the microgravity environment is 64% higher than the number of atoms on the ground. The increase in the BEC is most likely due to more efficient loading of atoms into the magnetic trap without the effects of gravity.

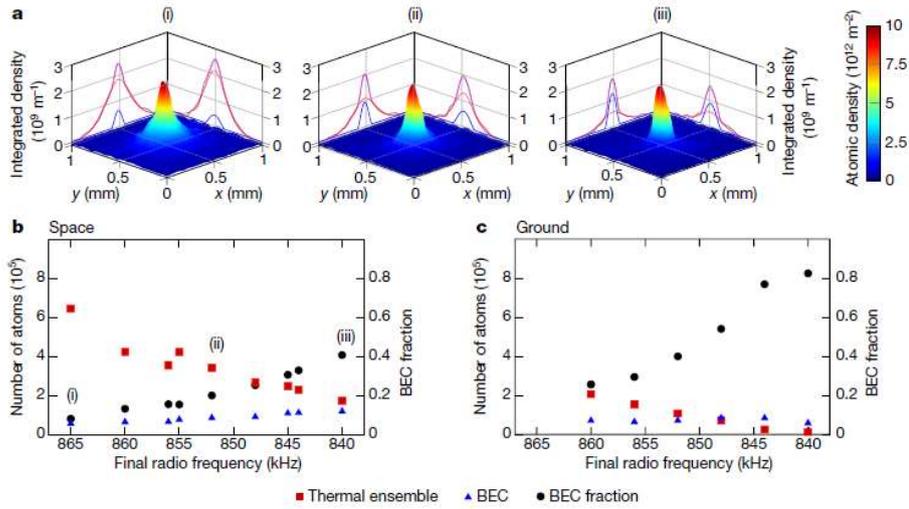

**Fig. 17 (a)** Phase transition to BEC in space and on the ground, controlled by the final radio frequency of forced evaporation. Spatial atomic density (color scale) and corresponding line integrals (solid grey lines), as well as Gaussian (red lines) and parabolic (blue lines) fits of the line integrals of the thermal and condensed atoms, respectively, and their sum (violet lines), for cases in space where 8% (i), 20% (ii), and 41% (iii) of the atoms are in the BEC state. **(b) and (c)** The number of magnetically trapped atoms in the thermal ensemble (red squares, left axis) is higher in space **(b)** than on the ground **(c)**, resulting in more atoms in the BEC (blue triangles, left axis) in space; for a comparable BEC fraction, there are 64% more atoms in the BEC in space than on the ground [52].

The transportation and shaping of BECs is a major goal of interferometry, with the key step of forming a compact BEC wave packet and transporting it from one position to another. In this experiment, the BEC was moved 0.8 mm from the chip surface. They studied the effect of the transportation process on the shape of the BEC and found that the BEC was excited after being transported. They also measured the oscillation of the centroid position after BEC excitation and compared the measured results with a theoretical model of kinetics [53]. The theoretical calculations included the process of oscillation, release, and evolution of the BEC in the trap potential until it was measured. The theoretical model also included the current-carrying conductor structure of the experimental device, and the result was obtained by solving the Gross–Pitaevskii equation in the Thomas-Fermi region [100]. The experimental results agreed with theoretical simulation results (Fig. 18 (a) and Fig. 19 (a)). The motion on the atom chip caused a complex oscillation of the BEC [101] shape, which illustrates the importance of phase-stabilized manipulation in atom chip experiments.

It can be seen from Fig. 17 that the background temperature (Gaussian shape, red line) of the released atoms was about 100 nK, and the BEC (Parabolic shape, blue line) was equivalent to a few nanokelvins of kinetic energy, which corresponds well with the theoretical calculation. Further reductions in temperature require an additional procedure, such as DKC, which is not possible to complete in a short-range rocket flight. Thus, the space station offers great potential for achieving picokelvin temperatures through DKC. The microgravity environment of the sounding rocket provides an ideal platform for comparing theoretical predictions with observations and testing various techniques. Therefore, the results of these series of experiments have important implications for future experiments, such as the NASA Cold Atomic Laboratory (CAL) on the International Space Station (ISS) and the NASA-DLR Bose-Einstein Condensate Cold Atomic Laboratory (BECCAL), which is currently in the planning stage [140].

Corgier et al. [53] systematically studied the theoretical and experimental design of deep cooling to picokelvin temperatures with atom chips in a microgravity environment. They hope to build a new generation of atomic interferometers that should be able to interfere for a few seconds. Their final design would enable controlled atomic transportation and hold time using a well-designed, final DKC-step coordination, where the BEC can be moved by millimeters to achieve picokelvin temperatures and slowly expand. The design predicts an optimum final expansion temperature of 2.2 pk [53].

It is difficult to achieve slow adiabatic transportation of a BEC from one position to another in experiments. Therefore, Corgier et al. [53] used the theory of fast transportation [102] and the shorten-to adiabaticity (STA) scheme [103] to design experiments, and through theoretical calculations they hope to achieve fast, non-adiabatic transportation of BECs under clear boundary conditions [104].

Fig. 18 (A) schematically shows a Z-shaped current coil in which two wires aligned along the y-axis are 16 mm long and the wire size in the x direction is 4 mm. These wires carry a direct current of $I_w$ = 5 A. The BEC atom was initially trapped ($t$ = 0) at a position $Z_i$ = 0.45 mm from the surface of the chip, directly below the origin of the axis. By changing the position of the minimum value of the z direction in the trap, the trapped atoms are transferred from $Z_i$ to the end position of $Z_f$ = 1.65 mm. Transportation of the BEC occurs over a total distance of $Z_f$ - $Z_i$ = 1.2 mm, which is much larger than the typical size (micron) of BECs.

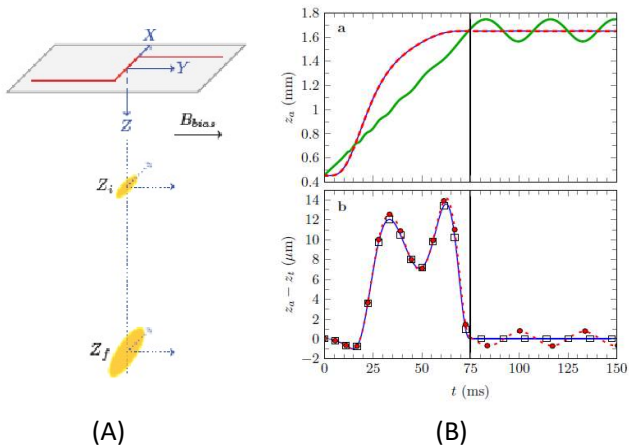

(A)           (B)

**Fig. 18 (A)** Schematic diagram of the chip configuration and of the displacement of the BEC. The Z-shaped wire is represented by the red line. **(B)** BEC position during and after the STA transport ramp. The upper plot **(a)** depicts the evolution of the expected position value $z_a$ of the BEC as a function of time in for a linear ramp (solid green), the harmonic trap (thin blue curve), and with a cubic term (dashed red line). The lower plot **(b)** shows the deviation from the trap position $z_f$ for better visibility of the STA ramp results [53].

They theoretically simulated the entire physical process, which includes the transformation of the BEC from the initial position $z_i$ to the final position $z_f$, further cooling using DKC during the holding time and preparation for the final stage of interference. The potential produced by the atom chip consists of two parts: harmonic and non-harmonic:

$$V_a(\vec{r},t) = \frac{1}{2}m\left[\omega_x^2(t)x^2 + \omega_y^2(t)y^2 + \omega_z^2(t)(z-z_t)^2\right] + \frac{1}{3}m\omega_z^2(t)\frac{(z-z_t)^3}{L_3(t)} \quad (17)$$

where $z_t$ denotes the minimum position of the trap potential in the z direction (the non-harmonic component mainly exists in the z direction) and $L_3(t)$ represents the characteristic length of the third-order non-harmonic term.

In the experiment, after evaporative cooling in the chip trap, the atomic gas became a BEC. Its wave function $\Psi(\vec{R},t)$ satisfies the Gross-Pitaevskii equation:

$$i\hbar\partial_t\Psi(\vec{R},t) = \left[-\frac{\hbar^2}{2m}\Delta_R + V_a(\vec{R},t) + gN|\Psi(\vec{R},t)|^2\right]\Psi(\vec{R},t) \quad (18)$$

where $m$ represent the atomic mass, $g = 4\pi\hbar^2 a_s/m$ depict the scattering amplitude, $a_s$ is the scattering length of the s-wave, $N$ means the number of atoms in the condensate, and the nonlinear term $gN|\Psi(R,t)|^2$ describes the average field two-body interaction energy. R in the Gross-Pitaevskii equation describes the fixed coordinate system: R ≡ (X, Y, Z). The trap potential is described as:

$$V_a(R,t) = \frac{1}{2}m\left[\omega_x^2(t)X^2 + \omega_y^2(t)Y^2 + 2\omega_{XY}(t)XY + \omega_z^2(t)(Z-z_t)^2\left(1 + \frac{2}{3}\frac{(Z-z_t)}{L_3(t)}\right)\right] \quad (19)$$

The centroid position of the BEC is defined as $R_a \equiv \langle\Psi(R,t)|R|\Psi(R,t)\rangle$. Under the simple harmonic potential approximation, the classical trajectory of BEC in the z direction can be determined by the following equation:

$$\ddot{z}_a(t) + \omega_z^2(z)(z_a(t) - z_t)\left(1 + \frac{z_a(t) - z_t}{L_3(t)}\right) = 0 \quad (20)$$

For the BEC to be stationary at the initial position $z_i$ and the final position $z_f$, the following conditions must be met:

$$z_a(0) = z_i \qquad \dot{z}_a(0) = 0 \qquad \ddot{z}_a(0) = 0 \quad (21)$$

and

$$z_a(t_f) = z_f \qquad \dot{z}_a(t_f) = 0 \qquad \ddot{z}_a(t_f) = 0 \quad (22)$$

The lowest position of the trap potential needs to be met:

$$z_t(0) = z_i \qquad \dot{z}_t(0) = 0 \qquad \ddot{z}_t(0) = 0 \quad (23)$$

and

$$z_t(t_f) = z_f \qquad \dot{z}_t(t_f) = 0 \qquad \ddot{z}_t(t_f) = 0 \quad (24)$$

Using Eqn. (20), the trajectory of the BEC could be found, as shown in Fig. 18(B)-a. The red line is the result of the STA transportation sequence, and the green line is the result of the linear sequence. The variable $t$ (=75 μs) represents the transportation termination time, the frequency of the corresponding potential in the x, y, and z directions is (12.5, 50, 49.5) Hz, and the hold time is 75 μs. Fig. 18(B)-b shows the deviation of the center of mass $z_a$ of the BEC from the valley $z_t$ of the trap potential during transportation. After 75 μs, $z_a$ and $z_t$ overlapped due to the effect STA of non-adiabatic transportation. Fig. 19(a) demonstrates the complete sequence of BEC including the transportation, holding, release, and delta-kick collimation. After the STA transportation sequence, the potential was turned off for a hold time $t_{hold}$, then the atoms free evolved for 100 ms ($t_{free}$, the free evolution time) and the DKC pulse was started, i.e., the trap potential was turned on for $δt$ = 4.48 ms. The corresponding trap potential had frequencies in the x, y, and z directions of (1, 7, 7.2) Hz. Calculations show that the hold time $t_{hold}$ is very sensitive to the final temperature. Taking the scale variation of the BEC in the x direction as an example, the blue curve in Fig. 19 (a) shows the dimensional variation of the released BEC in the x direction at three different times ($t_{free}$ = 29.4, 31.4, and 33.4 ms). The best option here is to consider the release time of 31.4 ms, then the atoms would be collimated. If a slightly lower hold time would immediately increase the volume of the BEC (dashed blue line in Fig. 19 (a)), a slightly higher hold time would result in an instantaneous compression of the BEC (dashed blue line in Fig. 19 (a)), followed by a quick expansion.

After 100 ms of free expansion of the BEC with the optimal choice of $t_{hold}$ ($t_{hold}$ = 31.4 ms), the mean-field interaction energy is almost completely in the form of kinetic energy. After applying the optical trap lens or DKC pulse (the trap has a frequency of (1.7, 1.7, 7.2) Hz, delta trap lens time= 4.84 ms), an expansion could be observed with an average velocity of approximately 25.3 μm/s in three spatial directions, which is equivalent to a temperature of 2.2 pk. The expansion velocities in x, y and z directions were 22.2 m/s (5.2 pK), 8.7 m/s (0.8 pK), and 8.2 m/s (0.7 pK), respectively. Fig. 19 (b) shows the full sequence with transportation, holding, release, and delta-kick collimation, leading to an average expansion rate at the picokelvin levels from all three directions. The distribution of Fig. 19 (c) can be obtained by optimizing the parameters. The white star marker shows the results of an expansion temperature of 2.2 pK.

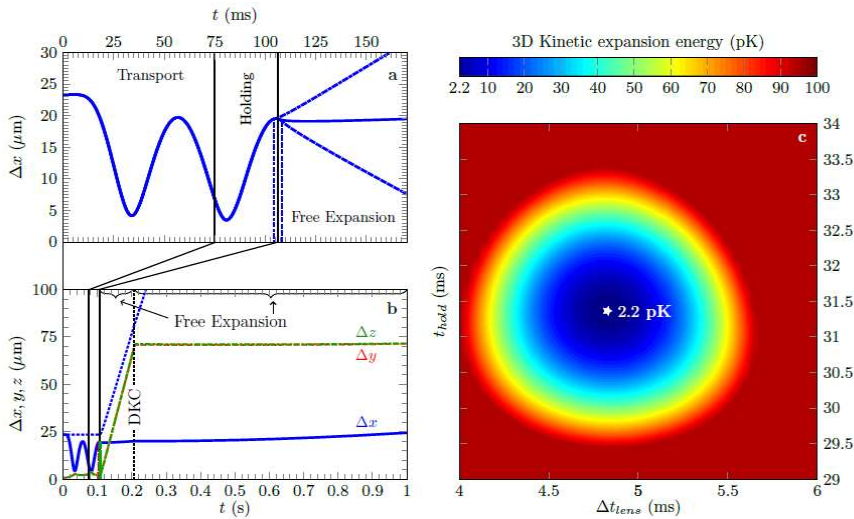

**Fig. 19** Transport, holding, release, and magnetic lensing of a BEC to picokelvin expansion velocities. **(a)** Effect of the release timing from the holding trap in the weak trapping direction x. The choices of 29.4 ms (dashed–dotted blue curve), 31.4 ms (solid blue curve), or 33.4 ms (dashed blue curve) illustrate different expansion behaviors (diverging, collimated, and focused, respectively). This timing has a little effect on the released size dynamics of the two strong axes y and z, not shown here for the sake of clarity. **(b)** Full sequence with transport, holding, release, and delta-kick collimation leading to an average expansion rate in the three spatial directions at picokelvin levels. **(c)** Optimal parameter search by scanning the holding and lens durations. The white star in (b) marks the optimal values leading to an expansion temperature of 2.2 pK [53].

**4.2 Experiments for CAL [54–55]**

There are currently two options for lengthy microgravity experiments: satellite-based experiments [106–108] and experiments on the ISS [109–114, 140]. Additionally, the Chinese Space Station is scheduled for 2022. Compared with satellite missions, experiments based on a space station are more attractive in terms of cost and duration because the instruments can be adjusted and upgraded. With these advantages, the NASA Jet Propulsion Laboratory proposed the establishment of the first Cold Atomic Laboratory (CAL) on the ISS to achieve ultracold Rb and K atomic gas with a temperature of 100 pK and then use the ultracold gas for scientific experiments.

The CAL is a multi-user platform that provides long-lasting picokelvin-scale temperature quantum gas under the microgravity conditions of space [54–55]. It uses a compact system based on atom chips to produce degenerate gas samples of $^{87}$Rb, $^{39}$K, and $^{41}$K ultracold mixtures. The CAL consists of three subsystems: scientific modules, electronic systems, and laser and optical systems. We describe only the scientific module here because it is core part of the CAL and is crucial to achieving picokelvin temperatures.

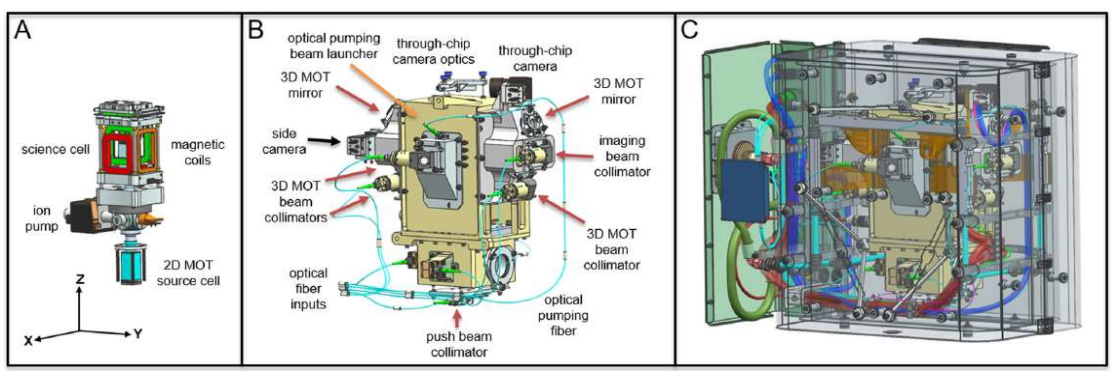

**Fig. 20 (A)** Physics package and magnetic coil assembly. MOT coils are shown in red, with the central axis of coils oriented along the x-axis. Transfer coils (in green) are vertically offset from the MOT coils. **(B)** Aluminum enclosure for the physics package securing the cameras and optical fiber collimators. Collimators and mirrors are emphasized to illustrate the beam geometry, with red/orange arrows indicating the effective directions of beam propagation. **(C)** The outside of this aluminum structure is bracketed to the water-cooling loop and further enclosed by the magnetic shields with feed-through connections from the other subsystems of the CAL instrument. Fully assembled, the dimensions are 46 cm × 30.5 cm × 58.5 cm, at a mass of 45 kg [55].

Fig. 20(A) shows the core of the scientific module [55], which consists of two glass absorption cells. The lower absorption cell is equipped with a source of Rb and K, and there are four permanent magnetic rods around the pool for the production of a 2D MOT, and the upper absorption cell is a 3D MOT. These two absorption cells are connected by a tube with an inner diameter of 0.75 mm to permit a pressure difference. The cold atoms (Rb and K) produced in the lower 2D MOT enter the upper 3D MOT chamber along the axis of the pinhole (Fig. 20(A)) with the help of a push light.

The CAL's scientific absorption cell is surrounded by ten rectangular magnetic coils (Fig. 20(B)) that are connected to water-cooling loops through eight thermal straps. These closed coils generate the necessary magnetic fields for MOT, magnetic transport, transfer to a chip trap, and Feshbach resonance detuning.

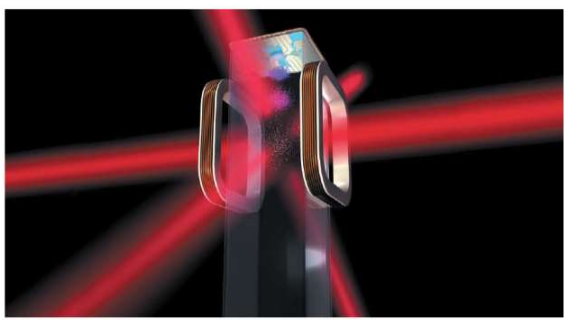
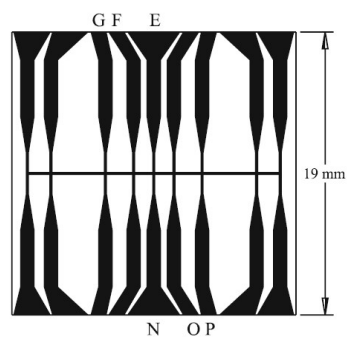

**Fig. 21** Instruments on NASA's Cold Atom Laboratory will cool atoms to near absolute zero (artist's concept).

**Fig. 22** Scale drawing of the CAL atom chip [56].

After the molasses phase, the cold atom cloud is loaded into the trap potential produced by the atom chip with the geometric structure shown in Fig. 22, which provides a z-trap, as shown in Fig. 18(a). The following is an analysis of how further cooling is conducted in this chip. There are two ways for deep cooling; one is the DKC process already used by the QUANTUS group and is described above, and the other is cooling by adiabatic expansion. The CAL plans to conduct both cooling experiments and compare them, and they have also carried out theoretical analysis of adiabatic expansion cooling [56]. Theoretical analysis shows that the z-trap of the atom chip can be used to obtain a three-dimensional effective temperature of 150 pK by adiabatic expansion cooling. The BEC lifetime in the CAL device is 10 s, which is the same as the experiment of MAIUS-1 in Section 4.1 (Fig. 18(a) introduces the same procedure in the experiment). However, the BEC trapped in the atom chip needs to move in the z direction in the actual experiment. The fundamental problem is to obtain the lowest temperature in a limited time without generating non-adiabatic excitation during the movement. If CAL initially produces $N = 10^4$ atoms in a potential trap and the condensation ratio is $N_0/N = 0.8$, these atoms will move by adiabatic expansion into a shallow trap with a frequency of 0.25 Hz. In this case, the final temperature of the system would reach 140 pK. The chemical potential of the condensate would be 28 pK, which is equivalent to the temperature of the condensate after closing the trap potential [61] to reach 8 pK. Because the atomic cloud has a 10% parametric excitation, the final temperature would increase to 155 pK and the condensing energy per atom would equal 9 pK. Then, the non-adiabatic excitation of the centroid motion would contribute 70 pK to the total energy produced by the ramp and contribute approximately 100 pK to the total energy produced by the uncompensated background gradient displacement. This gives us an ultralow temperature of 160 pK. Because the CAL studies also contain K atoms, the paper also estimates the cooling temperature of K atoms [56]. In the simple harmonic trap, the oscillation frequency of the K atoms is about 1.5 times that of Rb, so when the diffusion cooling is performed, the average frequency of the K atoms is given to be 0.37 Hz. With $10^4$ K atoms, this will result in a transition temperature of 360 pK.

**5. All-optical traps for deep cooling under microgravity conditions**

**5.1 Experiments for cold atoms in parabolic flight plane and Einstein elevator**

The atom chip uses magnetic confinement and cannot achieve a precise harmonic potential in three dimensions. Optical confinement overcomes this difficulty and has been used for deep cooling of an atomic gas under microgravity conditions. In 2011, Bouyer's team carried out an experiment of atomic cooling with optical Raman beams [48]. They employed a parabolic flight plane to cool atoms in a MOT plus Raman beams in microgravity and obtain images of the interference (Fig. 23 a–d). After the atoms were pre-cooled, they then applied a velocity-selective Raman light pulse carrying two counter propagating laser fields and cooled the atoms to a temperature of 300 nK in the longitudinal velocity distribution. The device is characterized by the usage of a fiber-optic laser in the communication band to form an all-fiber optical system, which makes the optical system more compact and integrated [117].

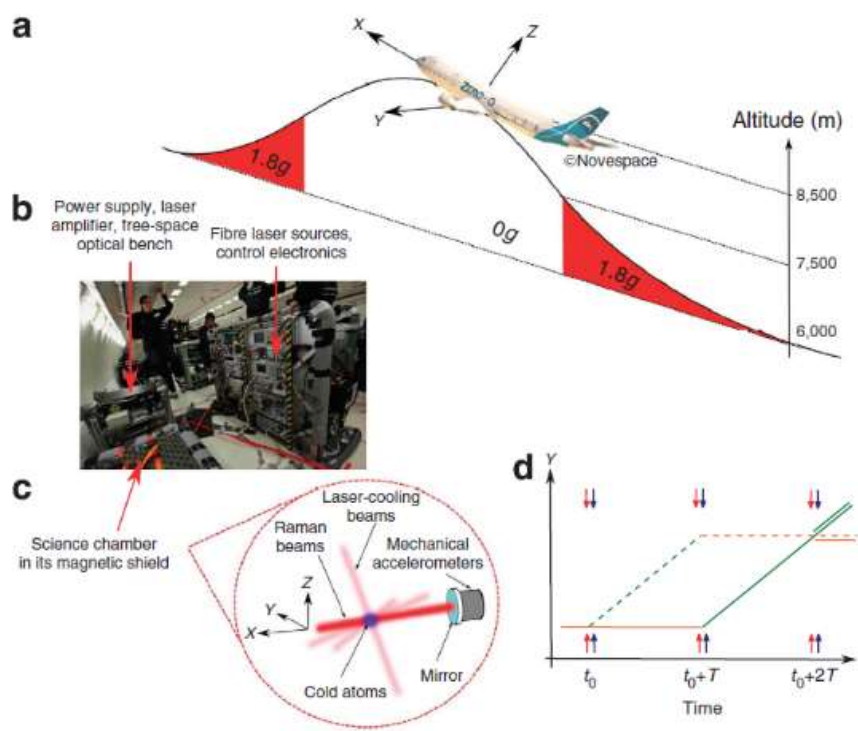

**Fig. 23** Description of the experiment in the parabolic flight plane. **(a)** The parabolic operation consists of a 20 s pull-up hypergravity (1.8 *g*) phase, a 22 s ballistic trajectory (0 *g*), and a 20 s pull-out 1.8 *g* phase. This operation is alternated with standard gravity (1 *g*) phases of about 2 min and carried out 31 times by the pilots during the flight. **(b)** Picture of the experiment in the plane during a 0 *g* phase. **(c)** Enlarged view of the science chamber where the atoms are laser cooled and then interrogated by the Raman laser beams (red) that are collinear to the Y axis and retroreflected by a mirror (blue). **(d)** Space-time diagram of the AI consisting of three successive light pulses that split, reflect, and recombine the two matter waves represented by the dashed and the solid lines. The blue and red arrows represent the two Raman laser beams[48].

Fig. 23 (a)–(b) shows the first use of a parabolic aircraft to investigate ultracold atoms and interferometers in a microgravity (~0 *g*) environment [48]. They use a plane called Novespace A300–0G [79] that can produce a 0 *g* environment repeatedly. During one flight on this plane, a 22-second parabolic ballistic flight (0 *g*) and a 2-min standard gravity flight (1 *g*) were available. With the parabolic flight, which can provide a microgravity environment of $10^{-2}$ *g*, the experiment can be repeated many times. During testing, they cooled $3 \times 10^7$ Rb atoms to 10 μK in 400 ms. Then, using a velocity-selective Raman pulse with two anti-propagation laser fields (Fig. 23 (c)), $10^6$ of Rb atoms with a temperature of 300 nK were obtained and an atomic interferometer was finally formed (Fig. 23 (d)). Experimental results show that the resolution of this matter wave interferometer based on microgravity is more than 300 times higher than the level of vibrations onboard the plane. Therefore, this experiment was a pioneer for future space missions such as the STE-QUEST proposed as ESA's candidate project of Cosmic Vision 2020–2022 [118].

To obtain colder atoms, they used a microgravity simulator (termed Einstein elevator) to cool atoms in an all-optical trap in microgravity [119]. The cooling scheme relied on the combination of three optical techniques. First, they used an enhanced grey molasses to create a reservoir of cold atoms. Second, they used a 1,550 nm optical dipole trap (ODT) to creates a transparency volume to store the atoms in the dipole trap without emission and reabsorption of near-resonant photons. Third, they generated a time-averaged optical potential by spatially modulating the ODT beam with an acousto-optic modulator (AOM). Both a high capture volume and fast evaporative cooling with a high collision rate for atoms in ODT were then achieved.

They realized BECs with all-optical trap in microgravity using a science chamber mounted on an Einstein elevator developed by the French company Symetrie (Fig. 24), which undergoes preprogrammed parabolic trajectories. The Einstein elevator can provide up to 400 ms of a 0 *g* environment every 13.5 s. The residual acceleration during the motion, as measured with a low-noise mechanical accelerometer (Colibrys SF3600), yielded a maximum amplitude of 1 m/s$^2$ and a root-mean-squared repeatability of 0.05 m/s$^2$ [Fig. 24(b)].

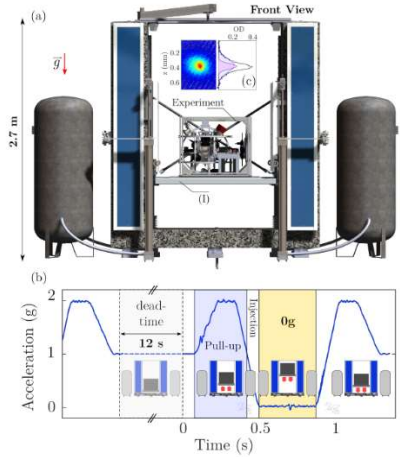 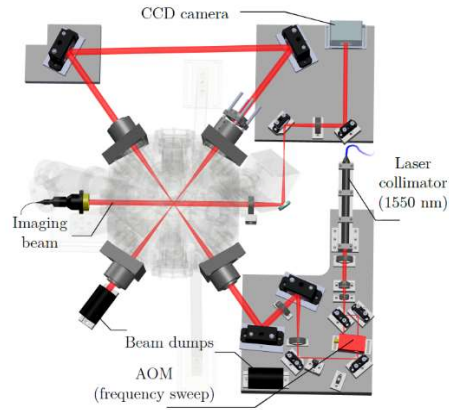

**Fig. 24 (a)** Schematic of the Einstein elevator. The payload (I) includes the science chamber, cooling beam optics, and imaging system. **(b)** Time sequence for the science chamber. **(c)** Absorption image of the BEC transition after a time-of-flight of 50 ms in 0 g[119].

**Fig. 25** Diagram of the optical trap composed of two crossed beams and the imaging system and the beam configuration of the dipole trap and probe [119].

Before the pull-up phase [Fig. 24(b)], they used the grey molasses [120] scheme to precool the atoms to 15 μK. Compared with red molasses, grey molasses has two advantages: shorter times for less expansion of the atoms clouds and more atoms can be loaded into the optical trap after the grey molasses phase. After that, the atoms were loaded into an optical trap composed of two crossed beams (Fig. 25), then evaporative cooling was performed to cool atoms to 400 nK. During the pull-up phase, evaporative cooling was maintained by different time sequences and evaporative cooling was completed by ramping down the ODP power until the end of the injection phase (the acceleration decreased from 1 g to 0 g). Using this protocol, a BEC was obtained 100 ms before the 0 g phase, with $5 \times 10^4$ atoms at a critical temperature of $T_c$ = 140 nK. At this point, the trap frequencies were approximately 100 Hz. At the beginning of the 0 g phase, the ODT power was decreased in 40 ms to reach the minimum value required to keep the atoms in the trap. At this stage, the BEC contained $4 \times 10^4$ atoms for a spatial expansion corresponding to 35 nK. The ODT power was 10 mW for an average trap frequency of 39 Hz. This experiment demonstrates the potential to cool atoms in microgravity with an all-optical trap.

**5.2 Two-stage cooling in an all-optical trap**

Evaporative cooling in an all-optical trap is an attractive technique for rapidly cooling atoms [120–124]. The principal limitation to evaporative cooling in the nanokelvin regime is the extended time required for thermalization due to the low speed of the atoms at ultralow temperatures. For instance, at a temperature of a few nanokelvin, there are only 2.6 collisions per second on average in a system of 3,000 Rb atoms in an optical trap. Such a low collision rate is not adequate for effective evaporative cooling. The useful cooling time is limited by inelastic collisions. To reach temperatures in the picokelvin regime within an acceptable time, the CAPR (Cold Atom Physics

Rack) system intended for the Chinese Space Station (expected to be launched in 2022) has adopted an all-optical trap with the TSC scheme，which was proposed by Peking University team [57–60] in 2013. In the first stage of this technique, two crossed laser beams with a narrow waist and high power are used to form an optical trap for runaway evaporation cooling. Then, in the second stage, atoms with low enough temperatures are loaded into the optical trap formed by a crossed laser beam with a wide beam diameter and weak power to conduct controllable decompression cooling to finally reach a temperature below 100 pK.

Compared with the traditional method of using a magnetic trap and microwave evaporation, crossed-beam dipole trap cooling has the advantages of a small potential trap, high trap frequency, high evaporation cooling rate, easy mode matching (in the process of optical molasses, where the first stage is evaporative cooling and the second is controllable decompressing cooling, the laser beam centers easily coincide), and convenient applications of quantum simulations in a 3D optical lattice. In contrast to the atom chip, the trap potential center overlaps both in the first stage of evaporative cooling and the second stage of controllable decompression cooling. Transportation of the atom cloud is unnecessary, which is convenient for operations in the ground state, and avoids heating and exciting the BEC. Therefore, the Chinese researchers proposed to perform TSC (also called the Two stage-Cross Beam Cooling (TSCBC) method) in the CAPR system[57–60]. Compared with the scheme in the magnetic trap [40], the expansion process and decompression cooling can be easily and effectively controlled in all-optical trap with TSC. The advantage of TSC is that the mechanic energy of the atoms in the trap can be easily be lowered by reducing the power and expending diameter of the laser and optima control of the time sequence in the second stage.

The feasibility of TSC scheme can be verified by theoretical simulation using the DSMC method. The simulation results show that the TSC scheme can effectively cool $^{87}$Rb, $^{133}$Cs, $^{39}$K, $^{40}$K and other atomic systems below the temperature of 100 pK under microgravity conditions. In 2018, the Peking University team verified the feasibility of the TSC program through ground experiments [60]. Using this method, $^{87}$Rb was cooled to a temperature of 3 nK. The following is a description of the theory of the Peking University team and the related ground experiments.

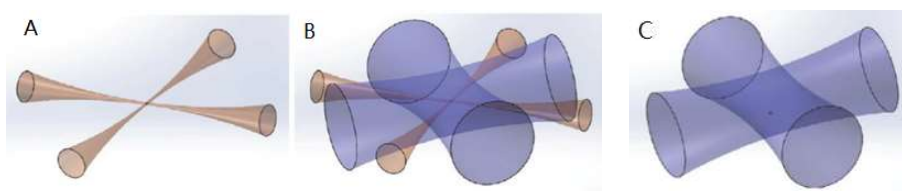

(a)

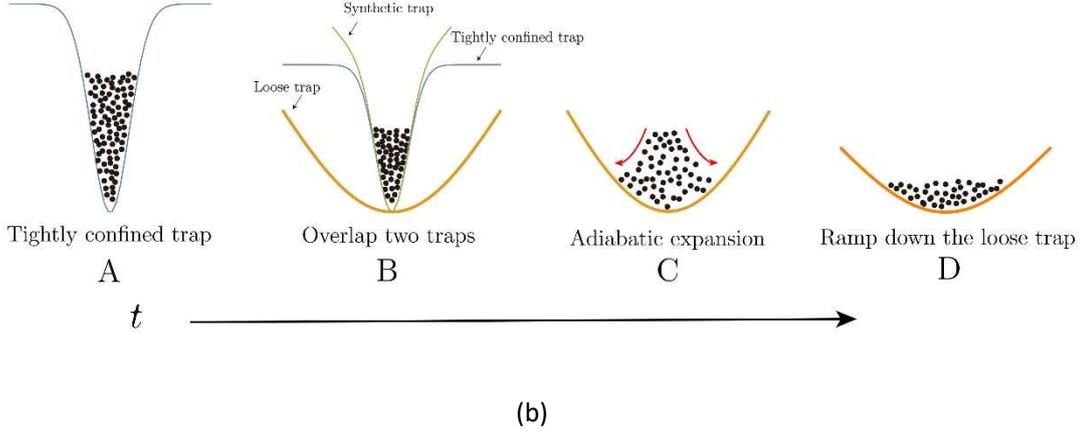

(b)

**Fig. 26 (a)** Proposed set-up of the multi-beam dipole trap and the two-stage approach to picokelvin temperatures in microgravity. The dark dots represent the atomic ensembles. **A**: Evaporative cooling in a very narrow crossed-beam dipole trap for 5 s. **B**: Overlapping the thin trap with a wider and weaker one. **C**: Decompression in the combined trap for 10 s. In the end, the narrow trap is shut down [57]. **(b)** Corresponding trap potential relating **(a)**, **A**: tightly confined trap potential from beginning of trap loading atoms corresponding **(a)-A, B**: the trap potential for two traps overlapped, corresponding to **(a)-B, C**: The atoms expand in the second trap potential, corresponding to (a)-C, D: the atoms adiabatically expand in the low trap potential, corresponding to **(a)-C**.

The literature [57–60] describes the basic principle and process of TSC to reduce the temperature of Rb and K atoms to the picokelvin region. In testing the TSC scheme, two mutually perpendicular intersecting beams were used (Fig. 26 (a)). The two identical laser beams recombined at their waists to form an optical dipole trap (ODT) (Fig. 26 (a) blue part). This potential field can be expressed as [123]:

$$U(r) = -\frac{3\pi c^2}{2\omega_0^3}\frac{\Gamma}{\Delta}\frac{2P}{\pi w_0^2} e^{-\frac{2z^2}{w_0^2}} \left( e^{-\frac{2x^2}{w_0^2}} + e^{-\frac{2y^2}{w_0^2}} \right), \tag{25}$$

where $c$ is the speed of light, $\omega_0$ depicts the center frequency of the atom transition, $\Gamma$ represents the spontaneous emission rate of the atom, $\Delta = \omega - \omega_0$ (the detuning frequency between ODT laser and the transition frequency of atoms), and $P$ is the power of the single beam laser. The geometric center of the crossing area can be approximated as a simple harmonic potential trap, as shown in Fig. 27 (a) (red part). Considering the influence of gravity, the trap potential of the crossed optical trap can be expressed as

$$U(r) = U_0 \left[ \frac{2z^2 + x^2 + z^2}{w_0^2} - 1 \right] - \alpha mgz \tag{26}$$

Where $g$ is the acceleration of gravity (under microgravity conditions of space station, preferably: $10^{-3} \sim 10^{-5} g$), $\alpha$ represents a coefficient ($from\ 10^{-5}\ to\ 1$), and the direction of gravity is along the z-axis. The trap depth should be

$$U_0 = \frac{3\pi c^2}{\omega_0^3} \frac{\Gamma}{\Delta} \frac{2P}{\pi w_0^2} \tag{27}$$

The potential energy of the simple harmonic trap at the center of the trap can be expressed as

$$U_c(x,y,z) = \frac{1}{2}m\left(\omega_x^2 x^2 + \omega_y^2 y^2 + \omega_z^2 z^2\right) \tag{28}$$

Where m is the atomic mass and the harmonic frequencies of the trap at three directions are:

$$\omega_x = \omega_y = \frac{\omega_z}{\sqrt{2}} = \sqrt{\frac{2U_0}{mw_0^2}} \tag{29}$$

Without gravity (i.e., $\alpha = 0$), the potential is a symmetrical trap (Fig. 27(b), blue dashed line). Under gravity conditions (i.e., $\alpha = 1$), the bottom of the original trap potential will be displaced in the direction of gravity (assuming z direction), $\Delta z = g/\omega_z^2$, forming a downwardly inclined trap potential (Fig. 27(b), blue solid line, sometimes called a tilted optical trap) in the direction of gravity, so that a gap is formed compared with the original trap potential (Fig. 27(b), blue dashed line). The trap depth at which an atom can be loaded is called the effective trap depth $U_{eff}$, which is lower than the potential $U_0$ of the original trap (blue dashed line). In a tilted optical trap, higher energy atoms escape in the direction of gravity. At the same time, compared with an optical trap with the same depth but no inclination (red dashed line), the trap frequency of an optical trap with inclination is significantly higher. In evaporative cooling, in order to keep the phase space density large enough, a significant number of atoms and low volume are needed to ensure the high density of the atom. Therefore, the effective trap depth $U_{eff}$ should not be a low value; it has a limit. Through analysis, we know that limitation of the effective trap depth $U_{eff}$ determines the lowest temperature of atoms on the ground (in the case without magnetic levitation). If the trap potential is less than the effective trap depth $U_{eff}$, the atoms will leak out of the trap with the action of gravity (Fig. 27 (c)).

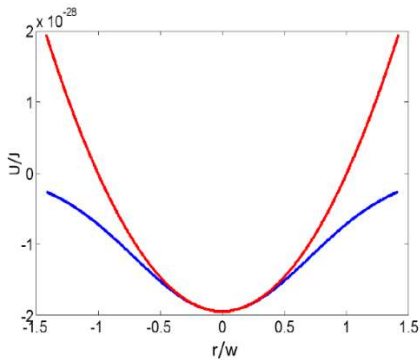

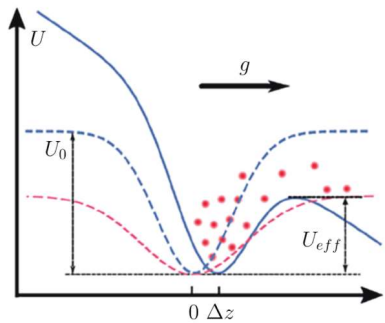

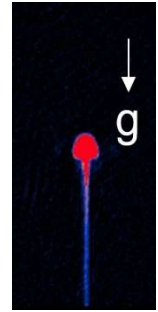

(a)                                                (b)                              (c)

**Fig. 27 (a)** Original optical dipole trap potential (blue line) and the ideal simple harmonic potential (red line). **(b)** Tilted potential (blue solid line) influenced by gravity with trap depth $U_{eff}$; Original optical dipole trap potential (blue dotted line) with trap depth $U_0$ and Untilted potential with same trap depth $U_{eff}$ (red dotted line). **(c)** Image of atoms leaking out of the trap under the action of gravity.

The direct simulation Monte Carlo (DSMC) method, adopted from the book *Molecular Gas Dynamics and the Direct Simulation of Gas Flows* by Graeme A. Bird (1998), can be used to calculate the average temperature after collision or expansion in an optical trap under different conditions of gravity (i.e., different α values in Eqn. (26)). The DSMC algorithm is essentially an improved method for molecular dynamics. The atomic trajectory and its collision characteristics still follow the thermodynamic equation. However, the method does not make a pair-by-pair decision about whether the atoms collide, but divides the entire calculation area *V* into a plurality of small calculation cells *ΔV*, then randomly extracts *ΔN* collision atoms in the calculation cell *ΔV* to collide. At this time, the computational complexity of the entire model is (V/ΔV)N, so the smaller *ΔV* is, the closer it is to the computational complexity of the molecular dynamics method. The size of *ΔV* is chosen to provide a reasonable value to improve program operation efficiency and to reduce redundant processing for collision-free atoms. A diagram of the DSMC simulation of ultracold atoms is shown in Fig. 28.

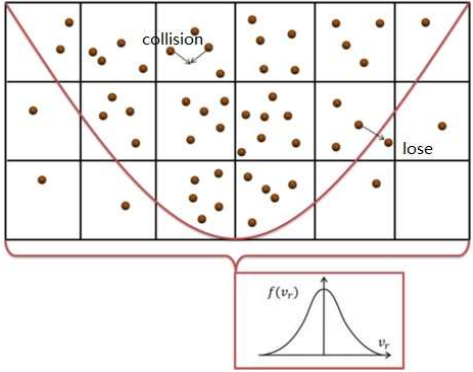

Fig. 28 Diagram of the algorithm in ultracold atomic system by DSMC simulation [66].

After approximating the Gaussian optical trap as a simple harmonic trap (Eqn. (26)), the motion of the atoms in the trap can be conveniently calculated. The initial state ($\vec{r}_0$, $\vec{v}_0$) of the atoms are set according to the scale of the trap and the initial temperature of the atomic gas, according to the simple harmonic equation of motion:

$$\frac{d^2 r}{dt^2} = -\omega_r r^2 \qquad (30)$$

The state of motion (r, v) of each atom after t time in the optical trap:

$$r = r_0 \cos\omega_r t + \frac{v_0}{\omega_0} \sin\omega_r t \tag{31}$$

$$v = v_0 \cos\omega_r t - r_0 \omega_r \sin\omega_r t, \tag{32}$$

where $r$ and $v$ are the position and velocity of atoms in three directions, $\omega_r$ is the frequency of the three directions of the trap potential determined by Eqn. (29). When dealing with collisions between atoms, because the entire calculation region $V$ can be divided into a plurality of small cells with the volume $\Delta V$, the number of collisions of atoms in a time $\Delta t$ is

$$\Delta N_c = \frac{n^2 \sigma v}{2} \Delta V \Delta t \tag{33}$$

In a randomly selected cell, there are $\Delta N_c$ pairs of atoms for collision, and a collision of two atoms with random exit angle is assumed. In a real system, in addition to elastic collisions, there are also two-body inelastic collisions, background gas collisions, and three-body composite collisions, all of which will decrease the number of atoms in the cell. The number of atomic losses per unit time in each cell can be estimated as

$$dn_{loss} = -K_2 n^2 - K_3 n^3 - \frac{n}{\tau_B} \tag{34}$$

where $n$ is the local density in the cell, $K_2$ represents the two-body loss coefficient, $K_3$ stands for the three-body loss coefficient, and $\tau_B$ depicts the lifetime of the atom in the trap potential. In the simulation, $K_2$, $K_3$, and $\tau_B$ are the data obtained from the experiment and applied in the program. By changing the light intensity of the optical trap, which is equivalent to changing the depth of the trap, we can calculate the position and velocity distributions of the atoms at a certain moment t, and finally determine the temperature of the system.

The typical 15 s TSC cooling process can be divided into two stages. The first stage belongs to a 5 s runaway evaporation process, and the second is a decompression cooling process for 10 s. Correspondingly, the tight-confining optical trap (Fig. 26 (a)) of the first stage are built by two orthogonal beams with 60 μm width, 1,064 nm wave length, and an initial power of 5–10 W. The wide-bonded optical trap of the second stage consists of two orthogonal beams with a 3 m waist and 30 mW of initial power (Fig. 26 (a)-C).

In the beginning of the first stage, the $10^7$ atoms that just emerged from the molasses phase with microkelvin temperatures are loaded into the tight-confining optical trap. The power of the trap beams is gradually reduced according to

$$P(t) = P_0 \times \left(1 + \frac{t}{\tau}\right)^\beta \tag{35}$$

where $P_0$ is the initial power of the laser, and $\tau$ and $\beta$ represent process parameters. Fig. 29 (a) is a plot of laser power as a function of time described by Eqn. (35). The optimal process parameters obtained by parameter scanning are $\tau$ = 0.03 Sec. and $\beta$ = 1.4. At the end of this phase, about $10^5$ $^{87}$Rb atoms with a temperature of 5.2 nK are trapped in the optical trap. The tight-confining optical trap still exists at the beginning of the second stage until the power is reduced to zero. Similar to the first stage, the power of the wide laser beams forming the wide-diameter bonded optical trap still have to ramp down, but the optimal process parameters come to $\tau' = 0.01$ s and $\beta' = 0.75$. Eventually, 3,000 $^{87}$Rb atoms with 7 pK were obtained at the end of the second stage (Fig. 29 (a)).

When TSC is carried out by using the cross laser potential well, the first stage of cooling (forced evaporation cooling, explained in detail in Section 2 with Eqns. (5) to (10)) lasts for 5 s. The duration of the second stage of cooling is generally 5–10 s, which is essentially a controlled decompression cooling process. Similar to the forced evaporation cooling process described in Section 2, the ordinal evaporation cooling is unrestricted, the difference is that forced evaporation cooling must be completed within a limited time. Similarly, the second stage of controlled decompression cooling, which is conditional adiabatic expansion cooling, must be finished in a finite time. Therefore, we need to consider how to change the potential of the optical trap in a limited time to improve the cooling effect. Eqn. (35) reveals the optimal values in the process of the two trap transformation.

The expansion process is realized by transferring atoms in first trap to the second trap. If the diameter of the second trap is much greater than the first, the volume increases substantially, while the potential depth will decrease; the atoms in the system will lose kinetic and potential energy simultaneously (Fig. 26 (b)-B,C). After the atomic expansion in the second trap, atoms will redistribute their kinetic and potential energy and begin to oscillate in the trap. Reducing the depth of the second optical trap would remove the atoms with higher energy from the trap; this procedure is the same function as evaporative cooling. The advantage of TSC is that, in the second stage, the mechanical energy of the atoms in the trap can be easily lowered by expending diameter and reducing the power of the laser and and optimal control of the time sequence.

Through DSMC simulation, the atomic temperature variation with time in the TSC scheme can be obtained (Fig. 29(c)), as well as change in the number of atoms (Fig. 29(b)). The curve in Fig. 29(c) shows that temperature of $^{87}$Rb atoms drops from microkelvin to nanokelvin within 5 s of the first stage of cooling. Then, within 10 s of the second stage of cooling, the temperature of $^{87}$Rb atoms continues to drop from nanokelvin to picokelvin.

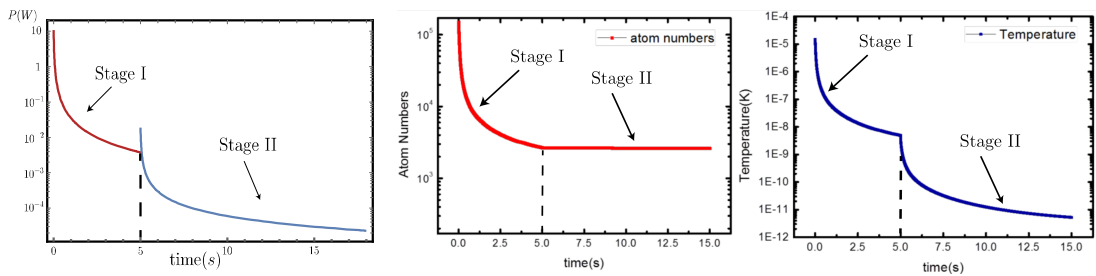

**Fig. 29** Evolution of **(a)** optical dipole trap laser power, **(b)** total number of atoms, **(c)** temperature versus time [57].

The DSMC simulates the distribution of each atom in the momentum space and position space before and after the second stage of cooling in the system, which can be compared in the same coordinate system, as shown in Fig. 30. Before the second stage of cooling, atoms are concentrated in position space and scattered in momentum space. After second stage of cooling, the atoms are scattered in position space and the momentum space is compressed, indicating that the secondary cooling reduces the temperature of the atoms. Nevertheless, its space density is not reduced, and the entropy in the system remains mostly unchanged.

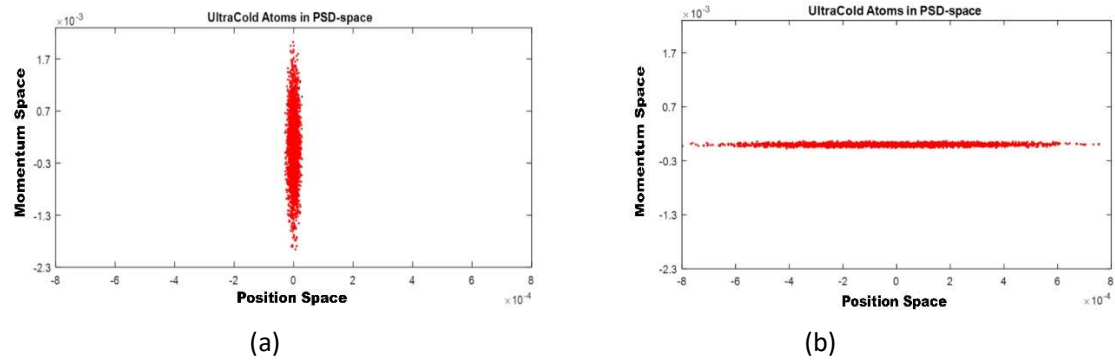

(a)  (b)

**Fig. 30** Position and momentum distribution of $^{87}$Rb atoms: before **(a)** and after **(b)** TSC cooling [66].

Because gravity produces a gap in the optical dipole trap (see Fig. (b)), the DSMC method can be used to calculate the final temperature of the system after TSC deep cooling under different gravity conditions $\alpha$ (Fig. 31). Fig. 31 shows that the final average temperature of the system is lowered with the reduction of gravitational acceleration. Therefore, obtaining an ideal microgravity environment is critical for obtaining picokelvin temperatures. For the TSC process, the shallow and wide laser beam cannot withstand a significant gravitational acceleration, resulting in a substantial number of atoms scattered in the direction of gravity (z direction). This process will lead to a decrease in the number of atoms and the density of phase space, so that the cooling cannot proceed normally, and the average temperature of the system is increased [59].

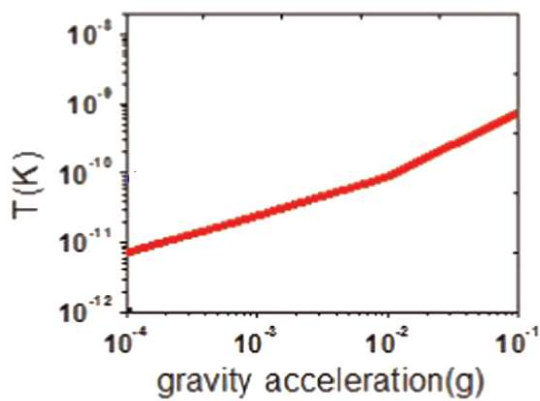

**Fig. 31** Relationship between final temperature and gravitational acceleration for TSC [59].

The above theoretical calculations result in an ideal state. In reality, the space station will have a variety of factors that are not necessarily considered on the ground that affects the deep cooling, such as the fluctuation of laser light intensity, the change in the gravity field, the vibration of the space station, and other factors. The influence of these factors on cooling is analyzed below.

Firstly, the effects of laser system fluctuations on the atoms in the trap are considered. A theoretical estimation is made to determine the range of laser power and frequency stability that should be controlled to ensure temperature stability of atoms on the order of 10 pK.

By Eqn. (26), the potential energy near the center of the dipole trap $U_{dip}$ can be approximated as:

$$U_{dip} = -\frac{3\pi c^2}{2\omega_0^3} \frac{\Gamma}{\Delta} \frac{P}{\pi w_0^2} \tag{36}$$

At extremely low temperatures, the temperature of an atom can be viewed as a statistic of kinetic $K$ and potential $U_{dip}$ energy in three dimensions, namely

$$\frac{3}{2} k_B T = K + U_{dip} \tag{37}.$$

Taking the derivative to obtain

$$\frac{3}{2} k_B \cdot \frac{\partial T}{\partial P} = \frac{\partial U_{dip}}{\partial P} = -\frac{3\pi c^2}{2\omega_0^3} \frac{\Gamma}{\omega_0 - \omega} \frac{1}{\pi w_0^2} \tag{38}.$$

With the related parameters for $^{87}$Rb atoms of $\Gamma$ = 6.0666 MHz and $\omega_0$ = 2π × 3.84 ×10$^{14}$ rad/s, the laser parameters of $\omega_1$ = 2π × 2.82 × 10$^{14}$ rad/s and $w$ = 60 μm, and the constants $k_B$ = 1.38 × 10$^{-23}$ J/K and $c$ = 3 × 10$^8$ m/s, the partial derivative of temperature to laser power can be obtained [59]:

$$\frac{\partial T}{\partial P} = 1.81 \times \left( \frac{pK}{10^{-3} mW} \right) \tag{39}.$$

It can be known from the above Eqn. (39) that the noise of laser power must be accurately controlled below 10$^{-3}$ mW so that the atomic temperature will be lowered to the order of 10 pK.

Additionally, the mechanical vibration of the space station will seriously affect the cooling process itself. According to the acceleration data of the ISS, the evaporation cooling process can be simulated using the DSMC method to quantitatively explore how mechanical vibrations affect the cooling process in different frequency domains. With the mechanical vibrations of the space station, the motion of the atoms in an optical dipole trap is a typical forced vibration and can be described by

$$\ddot{x} + \omega_0^2 x = A\omega_0^2 \sin(\omega t + \phi_0) \tag{40}$$

and

$$x(t) + dx = -\frac{A\omega_0^2}{\omega^2 - \omega_0^2} sin(\omega dt + \phi_0) + \frac{v_0}{\omega_0} \sin\omega_0 dt + x_0 \cos\omega_0 dt$$
$$+ \frac{A\omega_0}{\omega^2 - \omega_0^2} [\omega_0 \cos \omega_0 dt + \omega \sin \omega_0 dt] \tag{41}$$

$$v(t) + dv = -\frac{A\omega_0^2 \omega}{\omega^2 - \omega_0^2} \cos(\omega dt + \phi_0) + v_0 \cos \omega_0 dt - x_0 \omega_0 \sin \omega_0 dt$$
$$+ \frac{A\omega_0^2}{\omega^2 - \omega_0^2} [\omega \cos \omega_0 dt - \omega_0 \sin \omega_0 dt] \tag{42},$$

where $x$ and $v$ are the atomic displacement and velocity, and $\omega_0$ and $\omega$ are the trapping frequency and external vibration frequency, respectively. Before the TSC cooling, 7.5 × 10$^5$ 10 μK atoms are placed in an optical trap formed by a cross beam for 5 s. According to NASA's vibration data (red line in Fig. 32) for the ISS, the calculation can simulate the results of the final temperature (Fig. 33) and the final number of atoms (Fig. 34).

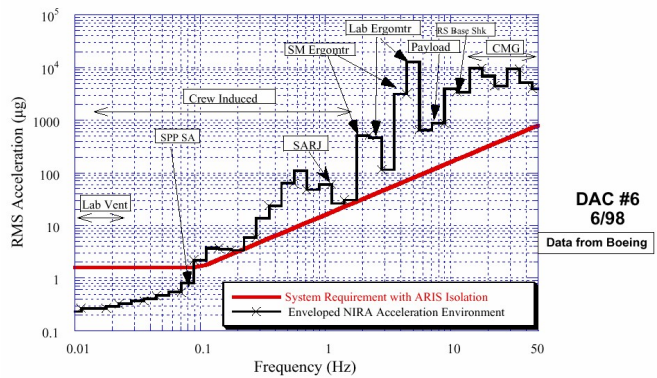

**Fig. 32** Acceleration data (Simulating with the red line) of International Space Station (ISS).

The simulation results show that in the lower frequency range (below about 50 Hz), the vibration causes serious atomic loss so that the cooling process stops. It can be found that if the vibration acceleration reduced by 10 times or 100 times, the serious atom loss is significantly suppressed, as shown in Fig. 33 (a) and (b). In the design of the actual system, the low-frequency vibration must be isolated to ensure that the final temperature enters the picokelvin range (Fig. 39 (a) and (b)).

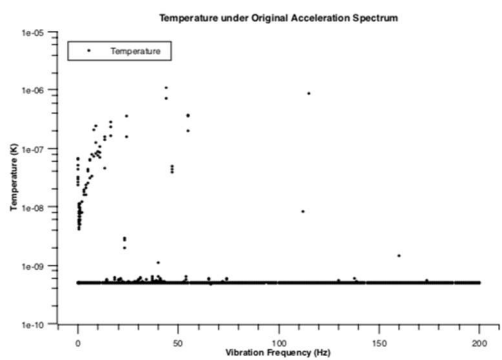
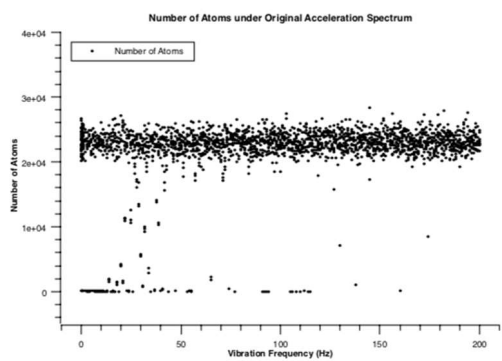

(a) final temperature  (b) final Number of Atoms

**Fig. 33** Vibration affection analysis according to the original acceleration spectrum of international space station: **(a)** Temperature and **(b)** Number of $^{87}$Rb atoms after 5 s evaporative cooling. Reprinted from arXiv:1909.00541

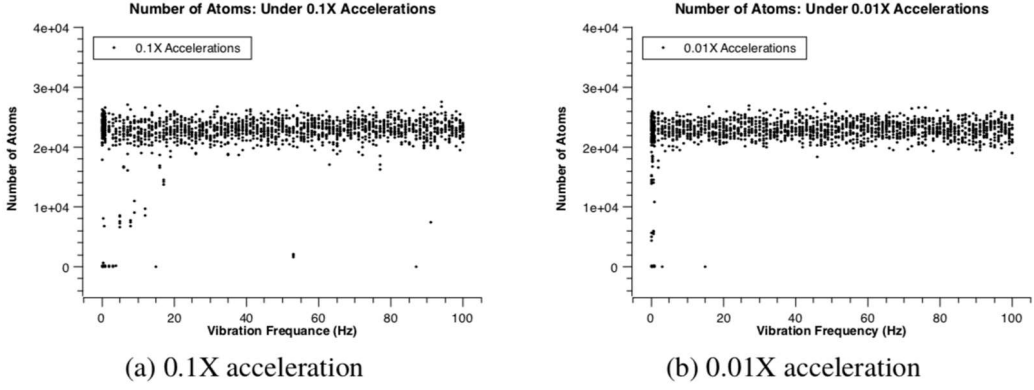

**Fig. 34** Vibrational effect analysis according to the reduced acceleration spectrum of the ISS: number of atoms after 5 s evaporative cooling. **(a)** For one-tenth of the original accelerations and **(b)** for one hundred times lower accelerations. Reprinted from arXiv:1909.00541

The above analysis shows that the microgravity environment is favorable for deep cooling, but the vibration of the space station itself and the fluctuation of light intensity will affect the results of the TSC cooling scheme. In order to obtain the temperature of atomic samples below 100 pK, corresponding preventive procedures such as vibration isolation and laser power stabilization must be taken.

In order to verify the TSC scheme, the team at Peking University tested the TSC scheme through the technique with an all-optical trap plus magnetic levitation [60], and they obtained the preliminary results of 3 nK. The experimental device for $^{87}$Rb Bose-Einstein Condensation they used is shown in Fig. 35 (a). After the pre-cooling process, including MOT, dark MOT, and optical molasses, 1 x 10$^7$ $^{87}$Rb atoms at 130 μK were obtained. The next step is the TSC cooling sequence shown in Fig. 35 (b). The atoms are loaded in a pair of laser beams with a diameter of 45 μm and 8 W of power. The evaporative cooling were performed, and at the end of the first stage, there were 10$^5$ $^{87}$Rb atoms condensed to form a BEC at a temperature of 90 nK and a 90% condensation ratio. The corresponding trap frequency is 70 Hz. During the second cooling stage, a magnetic field that balances gravity is turned on. This field is generated by two pairs of coils, one of which is an anti-Helmholtz coil that generates a gradient magnetic field suspending the $^{87}$Rb atoms. The other pair consists of Helmholtz coils is used to reduce the radial binding effect of the anti-Helmholtz coil so that the released atoms are more free in the radial direction. The electric field generated by Helmholtz and anti-Helmholtz coils in three-dimensional space can be expressed as

$$U_B(x,y,z) = \mu B_z' \sqrt{\frac{x^2}{4} + \frac{y^2}{4} + \left(z - \frac{B_z}{B_z'}\right)^2} \qquad (45),$$

where $\mu = \mu_B/2$ is the magnetic moment for the $|1,-1\rangle$ state $^{87}$Rb atoms and $\mu_B$ is the Bohr magneton. $B_z$ is the bias magnetic field that the Helmholtz coils produce, and $B_z'$ is the gradient magnetic field that the anti-Helmholtz coils produce.

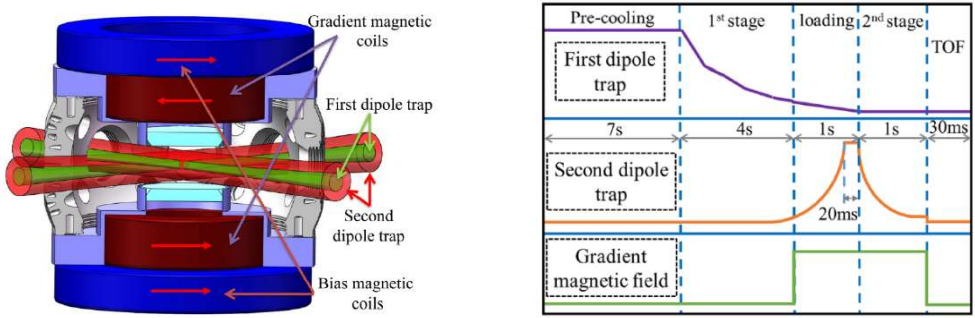

**Fig. 35 (a)** Schematic drawing of TSC experiment set. **(b)** The sequence chart of the TSC process [60].

Using Eqn. (45), one can estimate the radial binding frequency of the magnetic field coil:

$$\omega_x = \omega_y = \sqrt{\frac{\partial^2 U_B(x,y,z)/\partial x^2 \big|_{x=0}}{m}} = \sqrt{\frac{\mu B_z'^2}{4mB_z}} \quad (46).$$

This shows that to reduce the radial limit of an atom, one must increase the bias field $B_z$. In the experiment, $B_z$ = 70 G, and its equivalent gradient magnetic field is 30.5 G/cm, which is sufficient to balance the gravity of the $^{87}$Rb atoms in the ground state $|1,-1\rangle$.

In accordance with the timing sequence (Fig. 35(b)), the second pair of cross-optical dipole traps is adiabatically switched on while the gradient magnetic field is on. When the laser power reaches the maximum of 500 mW, the atoms originally in the first optical trap are converted to the second, whose diameter is 145 μm, and then held for 20 ms, while the first optical trap is almost turned off. The second stage of cooling, decompression cooling, is then conducted. In this process, the volume increases by 9 times that of the first trap, and the potential depth decreases significantly. After the atomic expansion in the second trap, the atoms lose their kinetic and potential energy simultaneously.

The laser power is reduced following Eqn. (35), and the optimal coefficients should be $\tau = 0.03$ and $\beta = 1$. Finally, the laser power of second optical trap is reduced to 3 mW, which is equal to optical potential depth of 1.2 E$_r$, and the final number of atoms is $4 \times 10^4$. Reducing the laser power reduces the depth of the second optical trap, which can remove the atoms with higher energy from the trap.

**Table 1**. relation between temperature and atomic speeds [66]

| T(nK) | $V_{3D}$(μm/ms) | $V_{1D}$(μm/ms) |
|---|---|---|
| 100 | 5.354 | 3.091 |
| 10 | 1.693 | 0.977 |
| 1 | 0.535 | 0.309 |
| 100 | 0.169 | 0.098 |
| 10 | 0.053 | 0.031 |
| 1 | 0.017 | 0.010 |

The final temperature can be measured using Eqn. (4). This requires maintaining the total number of atoms *N* so that the atoms will oscillate in the potential well. After measuring the frequency of the trap $\omega_{ho}$ and the background atoms around the condensation ($N$-$N_0$) in the trap, Eqn. (4) reveals the temperature of *T* = 3.5 nK for the background atoms, and the corresponding trap frequency $\omega_{ho}$ is approximately 5 Hz. Fig. 36 (a) shows the result of the oscillation of the BEC in the trap, which indicates the oscillation frequency $\omega_{ho}$ of the trap. Fig. 36(b) illustrates the relationship between the variation of the trap depth $U_{dip}$ and the trapping frequency $\omega_{ho}$ of the oscillation of the condensate in the trap. The black triangle data point is the measured value and the solid red line is given by Eqn. (29). The temperature of the atoms in the optical trap is lowered with the decrease in the trapping frequency $\omega_{ho}$ following the relation described by Eqn. (4). The final temperature in this experiment is limited by the noise of the magnetic coils for the levitation of the atoms. The results verify that TSC scheme in the all-optical trap is a suitable option.

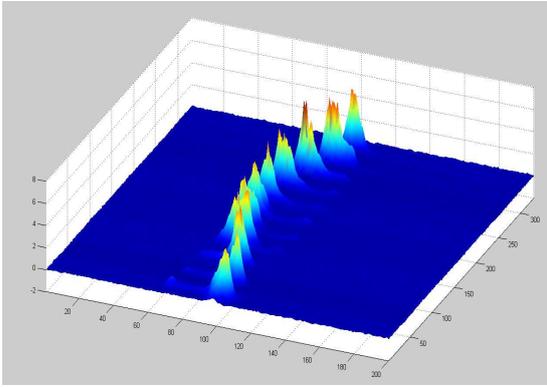 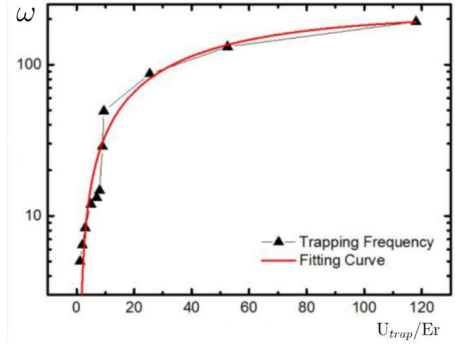

**Fig. 36 (a)** Oscillation of condensate in the optical trap; the temperature of atoms is around 3.5 nK. The atoms are cooled by TSC [66].

**Fig. 36 (b)** Relationship between optical trap depth and trapping frequency; the temperature of the atoms is proportional to the trapping frequency described by Eqn. (4) [66].

## 6 Summary and outlook

Obtaining picokelvin temperatures not only expand our understanding of the nature of the materials but also promote new technological applications. In terms of physical studies, Fig. 37 shows that there are different interactions between the atoms in the many body system at different temperatures [126]. The system appears to have a variety of phases in different energy scales (corresponding to temperature) because of the distinct regions of interactions, which determines magnetism, spin textures, and topological excitation, among other properties.

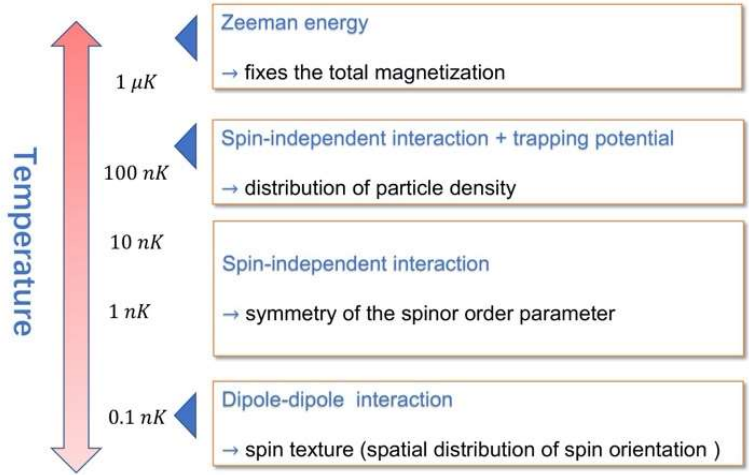

**Fig. 37** Energy (temperature) scales vs. interaction

In the quantum simulation experiment of an optical lattice (Fig. 38), if the temperature of the atom is low enough, the quantum tunneling between the lattice will be larger than the fluctuation caused by the temperature of the atom, and the accuracy of the experimental measurement will be greatly improved. Table 2 shows the hopping amplitude *t* (or tunneling flow) according to the Hubbard model in the optical lattice with different optical trap depths $U_0(E_r)$, the onsite-interaction energy *U*, and the tunneling flow *J* corresponding to the Heisenberg model. Thus, as the optical depth increases, the corresponding energy (expressed in terms of temperature) of these tunneling flows would reach the picokelvin order. If the actual temperature of the system can be reduced to the picokelvin region, the thermal fluctuation will not affect the interaction between the quantum tunnel and the field, which will greatly improve the experimental results.

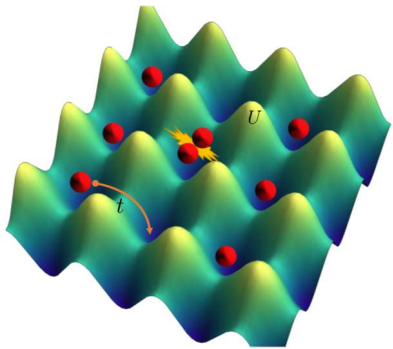

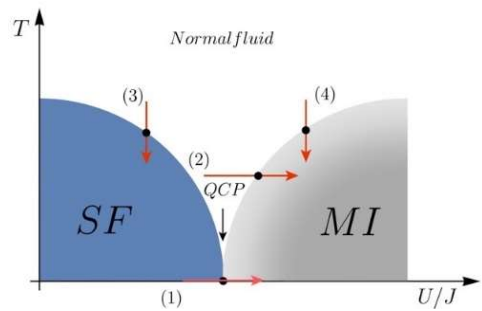

**Fig. 38** Atomic interactions in an optical lattice, tunneling between neighboring lattices (*t*) and interaction in one lattice (*U*).

**Fig. 39** Diagram of the finite *T* in an optical lattice. Phase transition routes (red solid line labeled 1 to 4) will undergo various features and follow distinct scaling laws.

In the case of phase transitions in an optical lattice with Bosons (Fig. 39), the quantum phase transition behaves with different characteristics at the finite temperature and absolute zero (*T* = 0

K) [26]. Fig. 39 shows a simplified scheme of the finite *T* phase diagram of Rb atoms in an optical lattice at mean lattice occupancy number *n* = 1. Strictly, the transition from a superfluid (SF) to a Mott insulator (MI) at the critical interaction strength $(U = J)_c$ in the optical lattice happens only at *T* = 0 K. For the region of $U/J < (U = J)_c$, the SF phase exists up to a critical temperature $T_c$, which decreases to zero at the quantum critical point (QCP), demonstrating a drastic change in the ground state of the system. Therefore, the transition routes in Fig. 39 (red solid line labelled 1 to 4) will undergo unique features and follow separate scaling laws, and the system transitions will reveal different features while the finite temperature changes from nanokelvin to picokelvin. The article [127] shows that the Higgs amplitude mode in an optical lattice emerges at typical low temperatures, and the peaks of amplitude of the Higgs mode will vanish while the temperature *T* is higher than the hopping amplitude *t* in Table 2.

**Table 2** Parameters with relative energy (temperature)

| Optical trap depth $U_0$ | 10 Er | 15 Er | 20 Er |
|---|---|---|---|
| *T* (tunneling) | 0.02 Er = 2.2 $K_B nK$ | 0.007 Er = 0.7 $K_B nK$ | 0.003 Er = 0.3 $K_B nK$ |
| *U* (interaction) | 0.28 Er = 27 $K_B nK$ | 0.38 Er = 37 $K_B nK$ | 0.47 Er = 46 $K_B nK$ |
| $J = \dfrac{4t^2}{U}$ (Tunneling in Heisenberg model) | 0.007 Er = 0.7 $K_B nK$ | $5.8 \times 10^{-4}$ Er = 0.056 $K_B nK$ | $6.6 \times 10^{-5}$ Er = 0.006 $K_B nK$ |

Note: $E_r$ is the photon recoil energy in the optical lattice, $K_B$ is Boltzmann constant, and K is Kelvin.

When the temperature of atoms decreases from nanokelvin to picokelvin, the wavelength of the matter waves further increases, improving the spatial coherence of the atom interference. Furthermore, because the accuracy of an atom interferometer is proportional to the square of the free-space evolution time, an atom interferometer at the lower temperatures of a microgravity environment will both improve coherence and measurement accuracy. This could become the basis for a next generation of high-precision basic physics applications. The interferometer could contribute to the search for dark energy [128], quantum testing of Einstein's equivalence principle [129–130], and long-baseline gravitational wave detection [131–132], helping to improve our understanding of general relativity and quantum mechanics. Additionally, the atomic clock based on ultracold atoms in microgravity [133] will provide high precision for various measurement and tests in fundamental physics.

An atomic interferometer at picokelvin temperatures could be expected to have an accuracy for measuring gravity with better precision than 10[-9] [134], which would be a powerful tool for Earth science, from measuring gravity at the Earth's surface and the spatiotemporal mass distribution under the surface. Beyond that, it could also monitor mass changes of icebergs [135–136] and volcanoes [137], as well as groundwater resources [138–139].

Thus, achieving temperatures in the picokelvin regime can not only stimulate new physics but also new technologies.

**Acknowledgments**

This work is supported by the National Natural Science Foundation of China (Grants Nos. 91736208, 11920101004, 61727819, 11934002) and the National Key Research and Development Program of China (Grant No. 2016YFA0301501) and the We acknowledge Daniel Kleppner for his helpful comments on the manuscript, we are also grateful to the discussions and comments with Jean Dalibard, Philippe Bouyer, Baptiste Battelier, Joerg Schmiedmayer, Tian Luan, Xiaoji Zhou, Wei Xiong, Bing Wang, Liang Liu.